\documentclass[floatfix,aps,pre,showpacs]{revtex4}

\usepackage{graphicx}
\usepackage{amsmath}
\begin{document}

\title{Computational core and fixed-point organisation in Boolean networks}
  \author{L. Correale} 
\affiliation{Politecnico di Torino, Corso Duca degli Abruzzi 24,
I-10129 Torino, Italy}

\affiliation{ISI Foundation, Viale Settimio Severo 65, Villa Gualino,
I-10133 Torino, Italy}

\author{M. Leone}
\affiliation{ISI Foundation, Viale Settimio Severo 65, Villa Gualino,
I-10133 Torino, Italy}

\author{A. Pagnani}
\affiliation{ISI Foundation, Viale Settimio Severo 65, Villa Gualino,
I-10133 Torino, Italy}

\author{M. Weigt}
\affiliation{ISI Foundation, Viale Settimio Severo 65, Villa Gualino,
I-10133 Torino, Italy}

\author{R. Zecchina}
\affiliation{International Centre for Theoretical Physics, Strada
Costiera 11, P.O. Box 586, I-34100 Trieste, Italy}

\date{\today}

\begin{abstract}
In this paper, we analyse large random Boolean networks in terms of
a constraint satisfaction problem. We first develop an algorithmic
scheme which allows to prune simple logical cascades and
under-determined variables, returning thereby the computational core of
the network. Second we apply the cavity method to analyse number and
organisation of fixed points. We find in particular a phase transition
between an easy and a complex regulatory phase, the latter one being
characterised by the existence of an exponential number of
macroscopically separated fixed-point clusters. The different
techniques developed are reinterpreted as algorithms for the analysis
of single Boolean networks, and they are applied to analysis and {\it
in silico} experiments on the gene-regulatory networks of baker's
yeast ({\em saccaromices cerevisi\ae}) and the segment-polarity genes
of the fruit-fly {\em drosophila melanogaster}.
\end{abstract}

\pacs{05.20.-y, 05.70.-a, 87.16.-b, 02.50.-r, 02.70.-c}

\maketitle

\section{Introduction}
One of the most surprising results of sequencing the human genome is
the small number of genes found: We have only about 25 000 genes,
compared to, {\it e.g.}, the about 19 000 genes of the simple worm
{\it C. elegans} \cite{Hgp}. This implies that the higher complexity
of the human organism cannot be understood in terms of the number of
genes. In addition, all different cell tissues -- humans have more
than 200 -- carry the full genetic information, but the corresponding
gene expression patterns, {\em i.e.} the set of genes which are
actually transcribed into mRNA and translated into proteins, are very
different from one tissue to another. In this sense, different cell
types can be understood as different robust {\it expression states} of
the full genetic information. Simple mono-cellular organisms, in
contrast, have generally only one expression state.

The key process hereby is {\it gene regulation}. The transcription of
a gene into mRNA is realised by a RNA polymerase. The action of this
molecular machine is enhanced or suppressed by the presence of
transcription factors (TF) binding to the DNA region upstream of the
gene. TFs are themselves proteins, encoded in other genes. This
defines a (directed) regulatory interaction from the gene coding for
the TF to the regulated gene. Only if the first one is expressed, the
TF is present in the cell, and can thus up- or down-regulate the gene
expression. The set of all regulatory interactions is called the {\it
gene-regulatory network} (GRN).

\begin{figure}[htb]
\vspace{0.2cm}
\begin{center}
\includegraphics[width=0.5\columnwidth]{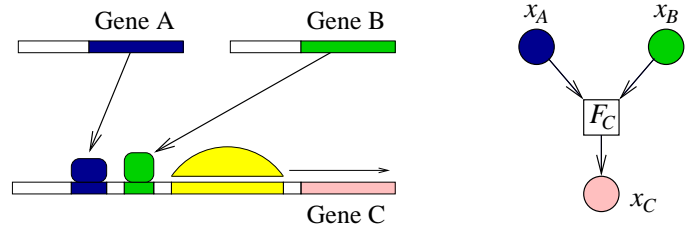}
\end{center}
\caption{Schematic representation of the regulation of a gene C by two
other genes A and B who code for transcription factors. These proteins
bind to the promoter region of gene C, regulating thereby the binding
of DNA polymerase II (the yellow semicircle) which is responsible for
the transcription process. The right-hand side of the figure
illustrates the Boolean model for this regulatory interaction: Genes
are represented by binary variables, and their combinatorial control
of the transcription of C is coded via a Boolean function.}
\label{fig:gene_regulation}
\end{figure}

In the last two decades a large wealth of new data about the
genome-wide organisation of gene-regulatory networks has been and is
still being collected \cite{LokWinz2000}. It has become clear that in
many cases biological functions cannot be identified at the level of
single or few genes and proteins. Constructing a detailed biochemical
model of an entire cell by analysing each gene and the nature of its
interactions with others one by one appears, however, to be an
intractable task. Even in the case of a relatively simple cell like
yeast, the number of genes is slightly above 6000.  In order to make
progress in the understanding of the combinatorial aspects of gene
regulation, it is thus crucial to model these systems on a {\it
coarse-grained level} \cite{Kaern}. Genetic interactions have to be
encoded in a discrete GRN: nodes of this network represent individual
genes, and directed connections their regulatory interactions. Such
models are expected to give valuable insight into the collective
large-scale behaviour of the gene regulatory mechanism, and - through
the understanding of these discrete models - it seems to be possible
to infer and study emergent cooperative phenomena in real biological
systems which cannot be understood at the level of single genes
\cite{Albert,Tang}. Numerical simulations of these model allows
further on efficient {\it in silico} experiments which may serve as a
valuable input to the design of real experiments. Statistical physics
provides important tools for the analysis of such complex systems, and
in turn finds in them new technical challenges.

In order to approach this problem, we need to choose a class of
discrete models which on one hand is detailed enough to schematically
encode the underlying biological processes, but on the other hand is
simple enough to allow for a large-scale quantitative analysis. We
therefore decided to restrict the present work to {\it Boolean
Networks} (BN) which compress the full information about the
transcription of a gene into a binary variable. This choice allows for
a pretty complete characterisation of the topological and functional
properties we are going to analyse below. Note however, that all
applied methods can be simply generalised to more detailed
descriptions, as long as the discrete nature of modelling is retained.

BN are dynamical models originally introduced by S. Kauffman in the
late 60s \cite{Kauf69}. Since Kauffman's seminal work, BN have been
used as abstract modelling schemes in many different fields, including
cell differentiation, immune response, evolution, and gene-regulatory
networks (for an introductory review see \cite{GCK} and references
therein). In recent days, BN have received a renewed attention as a
powerful scheme of data analysis, and modelling of high-throughput
genomics and proteomics experiments \cite{IlyaBook2005}.

The bottom-line of previous research is the description and
classification of different attractor types being present in BN under
deterministic parallel update dynamics
\cite{Kauf69,Kauf1,samuelsson}. Special attention was dedicated to
so-called critical BN \cite{Derrida} situated at the transition
between ordered and chaotic dynamics. We follow here a complementary
approach: our main goal is to study the statistical properties of {\em
fixed points} of general BN, by recasting the original dynamical
problem into a {\em constraint satisfaction problem}, and then using
different techniques borrowed from statistical mechanics of disordered
systems \cite{MPZ,Libro_Martin}. A parallel approach was used in
\cite{Lagomarsino}, and some of the results are already discussed in
\cite{letter}.

%We analyzed the organization of fixed points in random Boolean
%networks identifying the sudden emergence of a computational core,
%whose existence is a necessary (but not sufficient) condition for a
%globally complex phase where all fixed points are organized in an
%exponential number of macroscopically separated clusters. This
%phenomenon is found to be robust with respect to the choice of the
%Boolean functions, and missing only in networks where all boolena
%functions are of AND or OR type. In addition, the size of the complex
%regulatory phase grow is found to grow the higher the number $K$ of
%inputs to the Boolean functions is. As a starting testing ground, We
%implemented these techniques in two gene regulatory networks (yeast,
%and fruit fly) already known in literature.

The paper is organised in the following way. In Sec.~\ref{sec:model}
we introduce the model focusing on Kauffman models with $K=2$ inputs
per Boolean function. In Sec~\ref{sec:cc} we introduce the notion of
the {\em computational core}, and investigate two different
percolation phenomena related to it. In Sec.~\ref{sec:cavity} we apply
the cavity method and introduce different message passing algorithms
that allow to characterise in detail number and organisation of fixed
points. In Sec.~\ref{sec:K34} we extend the picture to the cases
$K=3,4$ and present a conjecture for the general $K \ge 2$ case. In
Sec.~\ref{sec:appl} we test our approach in two specific biological
examples: the GRN of yeast ({\em saccaromices cerevisi\ae}) and a
subnetwork of the fruit-fly {\em drosophila melanogaster}.
Sec.~\ref{sec:concl} is devoted to conclusions of the present work and
to an outlook on various open question.

\section{The model}
\label{sec:model}
In the simplest and most coarse-grained setting, the expression level
of one gene can be modelled by a Boolean variable: the gene-expression
level $x_i$ of gene $i$ can be either $0$ (gene $i$ is not expressed)
or $1$ (gene $i$ is expressed) \cite{Kauf1}. Although gene-expression
levels (given by mRNA and protein concentrations) are certainly not
Boolean variables, it turns out that a Boolean level of description is
able to capture many essential features of gene regulation and
moreover can cope with the often only qualitative nature of biological
knowledge \cite{ShDoZh}. This simplified view has become more adequate
since the recent progress of biotechnology has made fundamental
biological mechanisms experimentally accessible at a {\it genome-wide
scale}. An example are gene-expression experiments, where the activity
of a large number of genes is examined at the same time, using
DNA-chip techniques whose output needs to be digitalised before
further processing in a Boolean framework.

\begin{figure}[htb]
\vspace{0.2cm}
\begin{center}
\includegraphics[width=0.55\columnwidth]{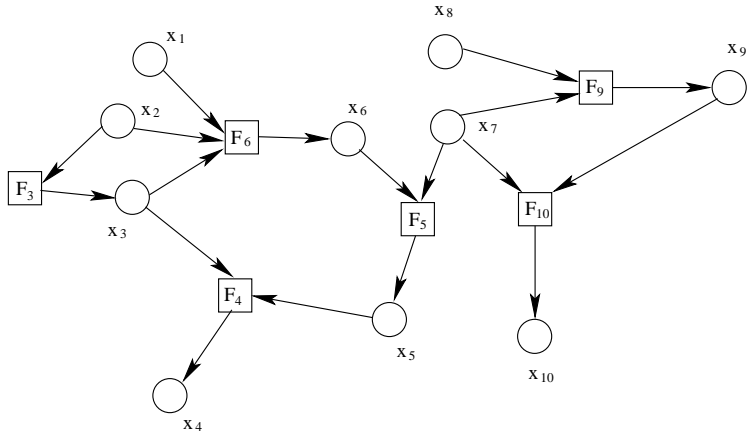}
\end{center}
\caption{Factor graph representation of a small Boolean network:
circles symbolise the variables, squares the Boolean functions. $x_1$
is an example for an external input variable, $x_4$ for a functional
variable, whereas $x_3$ stands for a transcription factor.}
\label{fig:fig1}
\end{figure}

Having $N$ genes (symbolised by circles in Fig.~\ref{fig:fig1}), the
global expression pattern can be described in terms of a
$N$-dimensional Boolean vector $\vec x = (x_1, \dots, x_N) \in
\{0,1\}^N$.  In this framework, gene-regulatory constraints are
modelled as Boolean functions (the squares in Fig.~\ref{fig:fig1}): the
expression level $x_a$ of gene $a$ is determined by a set of $K$
expression levels $x_{a_1}, \dots, x_{a_K}$ of the transcription
factors $a_1, \dots, a_K$:
\begin{equation}
\label{eq:function}
x_a = F_a(x_{a_1},...,x_{a_K})
\end{equation}
with $a\in A\subset \{1,...,N\}$ running over all regulated genes (see
\cite{Hwa} for a an interesting study on the different strategies
nature may adopt in order to implement Boolean regulatory rules). As
shown in Fig.~\ref{fig:fig1}, not all variables need to be controlled
by a Boolean function, i.e.~in general we have $|A|=M$ with $0\leq
M\leq N$. Note also that the actual value of the number $K$ of inputs
to $F_a$ may depend on $a$, even if the major part of this work will
concentrate on the case of constant $K$ for reasons of clarity.

The whole set of $M=\alpha N$ Boolean constraints together with the
$N$ variables define completely the network. The aim of this work is
to introduce new methods borrowed from statistical mechanics of
disordered systems. These methods are able to enumerate and classify
the ${\cal N}_{\mathrm{fp}}$ fixed points, {\em i.e.}~the set of all
vectors $\vec x$ fulfilling simultaneously {\it all} the $M$ Boolean
constraints of type Eq.~(\ref{eq:function}). With this aim we
introduce an Hamiltonian counting twice the number of not satisfied
Boolean constraints:
\begin{equation}
\label{eq:H}
{\cal H}(\vec x) = 2 \sum_{a\in A} x_a \oplus 
F_a(x_{a_1},...,x_{a_K}) = \sum_{a\in A} E_a(x_a,x_{a_1},...,x_{a_K}) \ .
\end{equation}
The symbol $\oplus$ stands for the logical XOR operation, {\em
i.e.}~each cost term contributes $0$ to the sum iff
Eq.~(\ref{eq:function}) is fulfilled, and $2$ otherwise. The prefactor
two is introduced for later convenience, and the notation $E_a$ is
simply an abbreviation which will be useful later on. The ${\cal
N}_{\mathrm{fp}}$ fixed points equal consequently the ground states of
Hamiltonian (\ref{eq:function}), provided that the minimum energy is
zero, {\em i.e.} that all constraints are simultaneously satisfied.
This embeds the problem of finding fixed points into the class of {\it
constraint-satisfaction problems}, which have been recently studied
extensively from the point of view of statistical physics, using
various techniques from spin-glass physics, in particular the replica
and cavity methods \cite{MPZ, Libro_Martin}, but also percolation-type
arguments.

In this work, we concentrate our attention to {\it random factor
graphs} subjected to two conditions:
\begin{itemize}
\item[({\em a})] Function nodes $F_a$ have fixed in-degree
$K$ and out-degree one. 
\item[({\em b})] Variables $x_a$ have in-degree at most one. This
means that all regulating variables are collected in one single
constraint $F_a$ (see Eq.~(\ref{eq:function})).
\end{itemize}
Setting $\alpha := M/N$, the degree distribution of variable nodes
approaches asymptotically
\begin{eqnarray}
\rho^{\mathrm{out}}(d_{\mathrm{out}}) &=& e^{-K \alpha }
\frac{(K\alpha)^{d_{\mathrm{out}}}}{d_{\mathrm{out}}!} \nonumber \\
\rho^{\mathrm{in}}(d_{\mathrm{in}}) &=& \alpha \delta_{d_{\mathrm{in}},1}
+ (1-\alpha)\delta_{d_{\mathrm{in}},0}
\label{eq:degdistr}
\end{eqnarray}
{\em i.e.}~the out-degree distribution is a Poissonian of mean
$K\alpha$, while in-degree distribution is bimodal. Since in- and
out-degrees are uncorrelated, the joint degree distribution
factorises: $\rho(d_{\mathrm{out}},d_{\mathrm{in}}) =
\rho^{\mathrm{out}}(d_{\mathrm{out}})
\rho^{\mathrm{in}}(d_{\mathrm{in}})$.  Random factor graphs are
obviously a drastic oversimplification for a true GRN, since the
available data show evidence for a scale-free $\rho^{\rm out}(d_{\rm
out})$ and a large but concentrated in-degree distribution of function
nodes more compatible with an exponential-like form
\cite{GuBoBoKe,Albert}. Nevertheless, this simplified model allows for
many detailed analytic predictions that can guide our comprehension in
more realistic and interesting cases, and is a well-controlled testing
ground for techniques borrowed from statistical mechanics. Such
techniques are not limited to random graphs and can be easily extended
to deal with more realistic cases.
\begin{figure}[htb]
\vspace{0.2cm}
\begin{center}
\includegraphics[width=0.55\columnwidth]{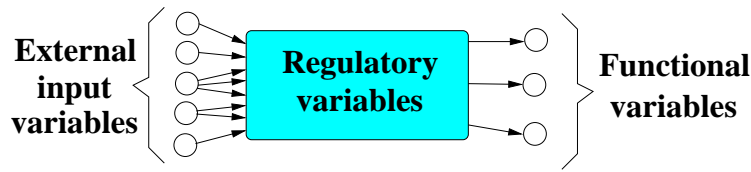}
\end{center}
\caption{Schematic representation of the three types of variables:
external input, functional, and regulatory.}
\label{fig:sandwich}
\end{figure}

In general, we can distinguish three sets of variables as
displayed in Fig.~\ref{fig:sandwich}:
\begin{itemize}
\item {\it External input variables:} There are $N-M = (1-\alpha)N$
variables which are not regulated by any function. They represent
external inputs to the network, as e.g.~chemical concentrations of
non-proteins. We do not consider these external inputs as dynamical
variables of the system. In the original definition of Kauffman
networks they are in fact included into the definition of the BN
itself: This definition works at $\alpha=1$ but allows for constant
functions, whose outputs are the analogues of our external input
variables.
\item {\it Regulatory variables} are all those variables which are
regulated and regulate. They can therefore be considered as
transcription factors.
\item {\it Functional variables:} There are $e^{-2\alpha} N$
variables which do not regulate any other function. These are to be
identified with functional proteins not involved in transcriptional
regulation, but acting instead as enzymes, membrane channels etc.
\end{itemize}
One has to note that in a random network there are some variables
which belong theoretically to both the external and the functional
variables because they are isolated. They contribute only trivially to
the behaviour of the network and can therefore be neglected in the
discussion.

\begin{table}[h]
\begin{tabular}{|cc|cc|cccc|cccccccc|cc|}
\hline $x_1$ &
$x_2$&0&1&$x_1$&$\overline{x_1}$&$x_2$&$\overline{x_2}$
&$\wedge$&&&&&&&$\vee$&$\oplus$&$\overline{\oplus}$\\
\hline 0 & 0 & 0 & 1 & 0 & 1 & 0 & 1 & 0 & 0 & 0 & 1 & 1 & 1 & 1 & 0 &
0 & 1\\ 0 & 1 & 0 & 1 & 0 & 1 & 1 & 0 & 0 & 0 & 1 & 0 & 1 & 1 & 0 & 1
& 1 & 0\\ 1 & 0 & 0 & 1 & 1 & 0 & 0 & 1 & 0 & 1 & 0 & 0 & 1 & 0 & 1 &
1 & 1 & 0\\ 1 & 1 & 0 & 1 & 1 & 0 & 1 & 0 & 1 & 0 & 0 & 0 & 0 & 1 & 1
& 1 & 0 & 1\\ \hline
\end{tabular}
\caption{Truth table for all 16 boolean functions of $K=2$ inputs.}
\label{tab:k2fun}
\end{table}

We now have to specify the functions acting on top of the random
topology defined so far.  There are $2^{2^K}=16$ Boolean functions,
which can be grouped into 4 classes \cite{Kauf2}:
\begin{itemize}
\item[(i)] The two constant functions. 
\item[(ii)] Four functions depending only on one of the two
inputs, {\em i.e.} $x_1,\overline x_1, x_2,\overline x_2$. 
\item[(iii)] {\it AND-OR class:} There are eight functions, which are
given by the logical AND or OR of the two input variables, or of their
negations. These functions are {\it canalising}. If, e.g., in the case
$F(x_1,x_2)=x_1 \wedge x_2$ the value of $x_1$ is set to zero, the
output is fixed to zero independently of the value of $x_2$.  It is
said that $x_1$ is a {\it canalising variable} of $F$ with the {\it
canalising value} zero.
\item[(iv)] {\it XOR class:} The last two functions are the XOR of the
two inputs, and its negation. These two functions are not canalising,
whatever input is changed, the output changes too.
\end{itemize}

%\begin{figure}[htb]
%\vspace{0.2cm}
%\begin{center}
%\includegraphics[width=0.5\columnwidth]{k2}
%\end{center}
%\caption{The two family of 2 inputs boolean functions: the canalizing 
%(AND-OR) and the not canalising one (XOR).}
%\label{fig:K2}
%\end{figure}

We keep in mind that the case of Boolean functions depending on
exactly $K=2$ input variables is not general in gene regulation:
genes, in particular in higher eukaryotes, are frequently regulated by
much more TFs. However, the number of Boolean functions of $K$
variables increases as $2^{2^K}$. A complete classification of Boolean
function becomes intractable already for relatively small $K\geq 4$.
For clarity we therefore concentrate first on true $K=2$-functions
only, {\em i.e.}~on those in the AND-OR class and the XOR class. The
organisation of the fixed points does not depend on the relative
appearance of the functions within each class, but only on the
relative appearance of the classes. We therefore require $xM$
functions to be in the XOR class, and the remaining $(1-x)M$ functions
to be of the AND-OR type, with $0\leq x\leq 1$ being a free model
parameter. In this sense, for $K=2$, the network ensemble is
completely defined by $\alpha$ and $x$. Extensions to higher
$K$-values are discussed below.

\section{The computational core of a Boolean network}
\label{sec:cc}

Before coming to the analysis of the fixed-point structure of random
Boolean networks, we present an algorithmic procedure to determine the
{\it computational core} (CC) of a Boolean network under a given
configuration of external variables. 

Many variables can be fixed simply by following logical cascades
originating in the external variables, and the corresponding Boolean
constraints are satisfied accordingly. Some other constraints can be fulfilled
because the included variables are otherwise unconstrained, see below
for a precise definition. However, a BN may contain a subset of
equations which cannot be solved in such a simple way on the basis of
local arguments only. We define the CC as the maximal subnet formed by
these Boolean functions. 

In order to determine this core, we prune iteratively all the function
nodes contained in the before-mentioned logical cascades, and all
those containing enough under-constrained variables. If no such
computational core exists, the determination of fixed points is just a
trivial task. For a non-trivial organisation of fixed points ({\it
e.g.} in clusters as found in the sections below) the existence of an
extensive computational core is a necessary, although not sufficient
condition.

\subsection{Propagation of external regulation (PER)}
\label{sec:per}

\begin{figure}[htb]
\vspace{0.2cm}
\begin{center}
\includegraphics[width=0.55\columnwidth]{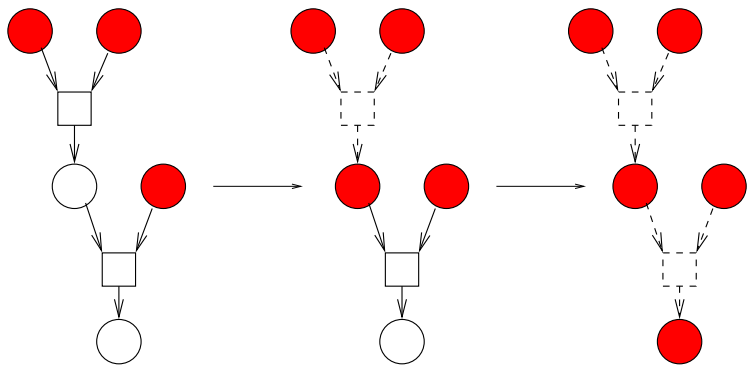}
\end{center}
\caption{PER: Both inputs to the upper left Boolean function are
  external, so also the output is directly fixed. Once this is done,
  also the inputs to the second function are fixed, and again the
  information can be propagated. All variables in this small sample
  graph are therefore determined by PER, no core exists.}
\label{per}
\end{figure}

Let us start our analysis by eliminating logical cascades which
propagate the external information into the network. If both inputs to
a Boolean function are external variables, then also its output is
fixed by the external condition, propagating thus the external
regulation to the output variable. As shown in Fig.~\ref{per}, this
argument can be iterated, and the external regulation can also be
propagated to variables depending not directly on external
variables. Functions with fixed output correspond to fulfilled Boolean
constraints, and they can be considered as being deleted from the
network. Note that, in case of a canalising function, the output can
already be determined by one input set to the correct value , {\em
i.e.}~the canalising one. We include also this propagation into the
PER analysis. The remaining part of the BN will be called the {\it PER
core}.

Note that, due to the input-dependent propagation via canalising
functions, the resulting PER core depends on the specific
configuration of the external variables, and is no longer a purely
topological property of the network. On the other hand, the PER core
size is expected to be self-averaging: almost all external input
configurations lead to the same PER core size.  Note also that a PER
core can exist only if the original BN contains feedback loops. The
latter can, however, be broken during the pruning process if
canalising functions are included. Directed percolation of the
underlying factor graph is thus a necessary but by no means sufficient
condition for the existence of a PER core.

Its size can be determined analytically using an iterative
probabilistic argument (which becomes exact in the large $N$ limit due
to both the uniqueness of the PER core for a given external condition,
and the locally tree-like structure of a random Boolean network). A
function has to be deleted from the network if either both inputs are
fixed (i.e. the inputs are either external or regulated by an already
eliminated function), or if only one of the inputs is fixed, but to
its canalising value. Denoting the fraction of all Boolean functions
which belong to the core by $\pi_{PER}$, we can thus write
\begin{equation}
1-\pi_{PER} = (1-\alpha\pi_{PER})^2 + \frac 12\ (1-x)\ 2\ 
(1-\alpha\pi_{PER})\ \alpha\pi_{PER}\ .
\end{equation}
The first term on the right-hand side describes the situation that
neither of the inputs depends on a function of the core, the second
term restricts to canalising functions (factor $1-x$) with exactly one
output depending on a surviving function. The prefactor 1/2 gives the
probability that the fixed input variable takes its canalising value.
We thus find for the survival probability
\begin{equation}
\pi_{PER} = 1-(1-\alpha\pi_{PER})^2 - (1-x)\
(1-\alpha\pi_{PER})\ \alpha\pi_{PER}\ .
\end{equation}
This equation has the obvious solution $\pi_{PER}=0$, corresponding to
the non-existence of a PER core. This solution is correct only for
small $\alpha$, at higher $\alpha$ a second solution exists. This
solution appears continuously, and its appearance can thus be
determined by linearising the above equation. We find easily the
$x$-dependent threshold
\begin{equation}
\alpha_{PER}(x) = \frac 1{1+x} \ ,
\end{equation}
which decreases from one for $x=0$ (pure AND/OR functions) to 1/2 for
$x=1$ (only XOR-class functions). At this line, a {\it percolation
transition} takes place. For $\alpha$-values below (or left) of the
transition line no extensive PER core exists, almost all variables can
be fixed by propagating the external condition via trivial logical
cascades. Fixed points of the network are thus trivial consequences of
external regulation. This is not longer true for larger densities of
the network. There an extensive part of the functions survives the
removal procedure, a PER core exists.

PER is related to the {\it percolation of information} studied in
\cite{GCK}. There, the relative evolution of two slightly different
initial configurations under synchronous update is studied. This can
analogously be studied by looking to the propagation of a bug, i.e., of a
single flipped variable $x_i$ in a fixed point configuration. The main
issue is the number of variables which are directly
influenced by this variable flip, or, more precisely, the expected
number $n$ of Boolean functions depending on $x_i$ which become
unsatisfied by the bug. On average, there are $2\alpha x$
non-canalising functions depending on a randomly selected $x_i$, and
$2\alpha (1-x)$ canalising ones. The first become always unsatisfied
if one variable is flipped, the canalising ones only if the other
input does not carry its canalising value (i.e. with probability 1/2).
We thus find
\begin{equation}
n = 2\alpha x + \alpha (1-x) = \alpha (1+x)\ .
\end{equation}
If this number $n$ is smaller than one, the bug is healed out with
high probability after just a few iterations, the system is
dynamically in its frozen phase. For $n>1$ the number of new bugs
increases exponentially, and the system diverges exponentially from
its initial fixed point: we are in the chaotic phase of the BN under
synchronous update. Marginal stability is held exactly at the PER core
transition line $\alpha_{PER}(x)$, establishing thus the connection
between the so-called critical networks and propagation of external
regulation.

\subsection{Leaf removal (LR)}
\label{sec:lr}

Whereas PER removes trivial logical cascades from the Boolean network,
LR is able to eliminate those functions which are trivially satisfiable
due to otherwise unconstrained variables \cite{XOR}. Such variables are {\it
leaves}, i.e. variables of total degree one. Depending on whether they
are inputs or outputs to the only function they are involved in, we
also speak of in- and out-leaves.

\begin{figure}[htb]
\vspace{0.2cm}
\begin{center}
\includegraphics[width=0.55\columnwidth]{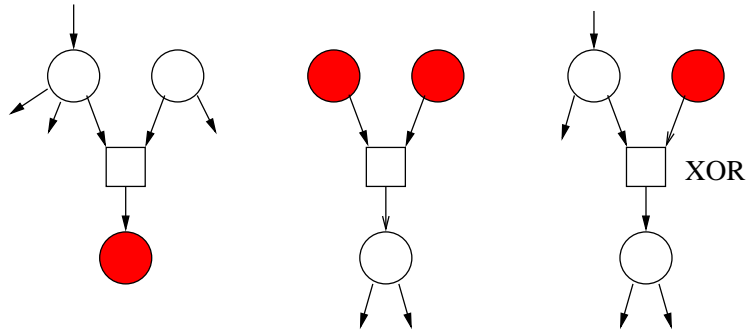}
\end{center}
\caption{LR: Cases, where a function can be pruned since it can be
satisfied easily due to otherwise unconstrained variables. Whereas the
first two cases, i.e. the out-leaf and the double in-leaf, are purely
topological and valid for any Boolean function, the rightmost figure
corresponds solely to the non-canalising XOR-type functions.}
\label{lr}
\end{figure}

In Fig.~\ref{lr} the cases of removable Boolean functions are
summarised. The first one is an out-leaf. Depending on how the inputs
are set, the output variable of the function can be set - without
influencing any other function. Similarly in the second case, where
both inputs to a function are leaves: If the output is fixed to some
value by consistency with the rest of the network, the inputs can be
adjusted easily (remember that no constant functions are present). For
the case of a XOR function, even one input being a leaf is sufficient
for removal: Whatever the value of the other two variables is, the
function can be satisfied by selecting the leaf-input accordingly. 

The removal procedure is not restricted to the leaves of the original
graph, but it can be iterated until no removable functions are
left. The remainder is called the {\it LR core}. Its size can again be
calculated on a random Boolean network using a simple iterative
scheme.  First we give a self-consistent equation for the probability,
that a variable becomes an out-leaf ($t_{\rm out}$) or an in-leaf
($t_{\rm in}$) at a certain point of the removal procedure. Let us
first consider out-leaves. They appear if all functions depending on
the variable are already removed. This probability reads
\begin{eqnarray}
t_{\rm out} &=& \sum_{d_1,d_2} e^{-2\alpha} \frac{ (2\alpha x)^{d_1}
(2\alpha(1-x))^{d_2} }{d_1! \ d_2!} 
\left[1-(1-t_{\rm out})(1-t_{\rm in})  \right]^{d_1} t_{\rm out}^{d_2} 
\nonumber\\
&=& \exp\{ -2\alpha (1-t_{\rm out}) (1-x t_{\rm in}) \}\ ,
\label{eq:to}
\end{eqnarray}
where $d_1$ is summing over the XOR-type functions, $d_2$ over the
canalising ones. We have further on taken into account that a XOR-type
function can be removed when either the other input or the output are
a leaf, whereas an AND/OR-type function is removable only iff the
output is a leaf (the case of two in-leaves is obviously excluded
since one of the inputs is the fixed one in this consideration). For
an in-leaf to appear, also the input function (if existing) has to be
removed, leading to the slightly modified expression
\begin{eqnarray}
t_{\rm in} &=& (1-\alpha+\alpha x (1-(1-t_{\rm in})^2) + \alpha (1-x) 
t_{\rm in}^2 )\ t_{\rm out} \nonumber\\
&=& (1-\alpha x (1-t_{\rm in})^2 - \alpha (1-x) (1-t_{\rm in}^2) )\ 
t_{\rm out}\ .
\label{eq:ti}
\end{eqnarray}
These equations close in $t_{\rm out}$ and $t_{\rm i}$, and allow us
to write the survival probability for an arbitrary Boolean function in
the network:
\begin{equation}
\pi_{LR} = x (1-t_{\rm out})(1-t_{\rm in})^2 + (1-x) (1-t_{\rm out})
(1-t_{\rm in}^2)
\end{equation}
where the first term accounts for the fact that an XOR-type function
survives if and only if none of its in- or outputs becomes a leaf at a
certain point, whereas an AND/OR-type function survives if neither the
output becomes a leaf nor both inputs are simultaneously leaves,
cf.~Fig.~\ref{lr}. The remaining fraction of XOR-class functions can
be evaluated as
\begin{eqnarray}
x_{LR} &=& \frac{x (1-t_{\rm out})(1-t_{\rm in})^2}{x (1-t_{\rm out})
(1-t_{\rm in})^2 + (1-x)
(1-t_{\rm out})(1-t_{\rm in}^2)}\nonumber\\
&=& \frac x{x+(1-x)\frac{1-t_{\rm in}^2}{(1-t_{\rm in})^2}}
\end{eqnarray}
which is smaller than $x$ because $1-t_{\rm in}^2>(1-t_{\rm
in})^2$. This observation reflects the fact that LR is more efficient
on non-canalising functions due to the rightmost removal case in
Fig.~\ref{lr}.

But does an extensive LR core exist at all, i.e.~is $\pi_{LR}$ larger
than zero? The obvious solution $t_{\rm out}=t_{\rm in}=1$ of
Eqs.~(\ref{eq:to},\ref{eq:ti}) -- every variable becomes a leaf at a
certain point -- leads to $\pi_{LR}=0$, i.e.~predicts the
non-existence of a core. For high $\alpha$, we find a non-trivial
solution which appears discontinuously at a threshold $\alpha_{LR}(x)$
which changes monotonously from $\alpha_{LR} (0) = 1/2$ to
$\alpha(1)=0.8839$. The discontinuous jump in the core size is maximal
for the pure XOR-type model ($x=1$): the value of $\pi_{LR}$ changes
from zero to about $0.386$.

To determine the threshold for arbitrary $x$, we observe that
Eq.~(\ref{eq:ti}) can be used to eliminate one of $t_{\rm out}$ and
$t_{\rm in}$. For the following, it is more practically to write
$t_{\rm in}$ as a function of $t_{\rm out}$. We find
\begin{equation}
t_{\rm in} = \frac{1-2\alpha x t_{\rm out}}{2\alpha (1-2x) t_{\rm out}} 
- \frac 1{2\alpha (1- 2x) t_{\rm out}} \sqrt{(1-2\alpha x t_{\rm out})^2 
- 4(1-\alpha)\alpha
(1-2x)t_{\rm out}^2 } =: T_{\rm in}(t_{\rm out},\alpha)\ ,
\end{equation}
where the sign in the solution of the quadratic equation was chosen to
guarantee that $t_{\rm in}\in [0,1]$ for $t_{\rm out}\in
[0,1]$. Plugging this into Eq.~(\ref{eq:to}) we obtain one single
equation
\begin{equation}
t_{\rm out} = \exp\{ -2\alpha (1-t_{\rm out}) [1-x T_{\rm in}
(t_{\rm out},\alpha)] \} =: T_{\rm out}(t_{\rm out},\alpha)
\label{eq:to_tr}
\end{equation}
for $t_{\rm out}$ as a function of $\alpha$. At the LR transition
point, two solution branches different from one appear
discontinuously, directly at the transition point with infinite slope
in $\alpha$. The lower one is the physical solution $t_{\rm
out}(\alpha)$. Let us denote $\alpha(t_{\rm out})$ its inverse
function. Plugging it into Eq.~(\ref{eq:to_tr}), the latter becomes
identically fulfilled. Deriving both sides with respect to $t_{\rm
out}$, we thus find
\begin{equation}
1 = \frac d{dt_{\rm out}} T_{\rm out}(t_{\rm out},\alpha(t_{\rm out})) 
= \frac {\partial
T_{\rm out}(t_{\rm out},\alpha)}{\partial t_{\rm out}}\ +\ \frac {\partial
T_{\rm out}(t_{\rm out},\alpha)}{\partial \alpha}\ \alpha'(t_{\rm out})\ .
\end{equation}
Directly at the LR transition, the derivative of $\alpha(t_{\rm out})$
vanishes (due to the infinite slope of its inverse function $t_{\rm
out}(\alpha)$). The partial derivative with respect to $t_{\rm out}$
can be evaluated, and we find that the transition line
$\alpha_{LR}(x)$ is given self-consistently by the equations
\begin{eqnarray}
t_{\rm out} &=& \exp\{ -2\alpha_{LR} (1-t_{\rm out}) 
[1-x T_{\rm in}(t_{\rm out},\alpha_{LR})]
\}\nonumber\\ 1 &=& 2 \ \alpha_{LR}\ t_{\rm out}\ \left[1-x\
T_{\rm in}(t_{\rm out},\alpha_{LR})+x(1-t_{\rm out})
\partial_{t_{\rm out}} T_{\rm in}(t_{\rm out},\alpha_{LR})
\right] \ .
\end{eqnarray}
The result is summarised in Fig.~\ref{perc}.

\subsection{Combining LR and PER}
\label{sec:lr_per}

\begin{figure}[htb]
\vspace{0.4cm}
\begin{center}
\includegraphics[width=0.65\columnwidth]{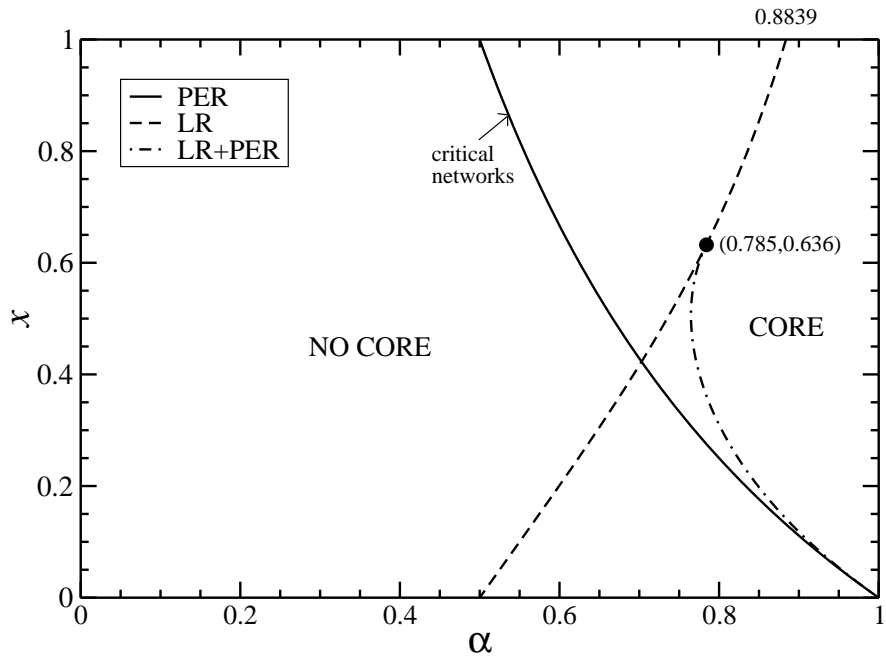}
\end{center}
\caption{Phase diagram for the percolation of the computational
core. We give the transition lines for PER, LR and the combination of
both. Left of the transition lines, no core exists, right of the
lines, a non zero core exists. The lines of LR and of LR+PER merge in
the tricritical point (0.785,0.636), below this point the core
appearance is continuous, above discontinuous.}
\label{perc}
\end{figure}

The two removal procedures can be combined to obtain the computational
core of a network (for a given configuration of the external
variables, cf. the discussion at the beginning of Sec.~\ref{sec:per}).
We first have to iterate the LR procedure to obtain the LR core, and
then to apply PER in order to understand which part of the LR core is
trivially fixed by propagating the external information via logical
cascades. Note that we cannot simply change the order of the two graph
reduction processes: whereas the argument for pruning Boolean
functions depending on one or two in-leaves relies on the fact that
these are not otherwise constrained, PER propagates constraints and
fixes at least some of the in-leaves of the PER core to non-arbitrary
values.

An immediate consequence of the combination of both procedures is that
the region showing a core is shrinking in the phase diagram to a part
of the intersection of the two regions with PER core and with LR core.
On random Boolean networks a complete analytical determination of the
percolation line and the size of the computational core is possible,
the technical details are explained in App.~\ref{app:lr+per}. Here
we will only review the main results:

\begin{itemize}
\item For $x<0.636$, a continuous emergence of the computational core
can be observed at some $\alpha_{CC}(x)$, cf.~the dash-dotted line in
Fig.~\ref{perc}. This line is situated deep inside the region where a
LR core exists, and approaches the PER transition line if we approach
$x=0$, i.e. the purely canalising case, from above.
\item For $x>0.636$, the existence of a LR core implies automatically
also the existence of a non-trivial computational core, we consequently
find $\alpha_{CC}(x)=\alpha_{LR}(x)$. The transition is discontinuous,
i.e. the size of the core jumps at the transition from zero to a
finite fraction of the full Boolean network.
\item Both regimes are separated by a tricritical point at
$(\alpha,x)=(0.785,0.636)$.
\end{itemize}

As discussed above, the determination of fixed points in the region
without a computational core is an easy computational task. Note that
in particular the critical networks studied in the context of
synchronous update dynamics are completely located within this easy
phase. The situation can become more complicated in the presence of a
computational core, since simple local methods to construct fixed
points (based on extensions of the discussed removal procedures) fail
to determine all variables. The percolation arguments presented so far
are, however, not sufficient to show that structure and organisation
of fixed points really changes as soon as an extensive computational
core of the full network exists. Answering to this question is the main
goal of the following two sections.

\section{The Cavity Approach}
\label{sec:cavity}
In this section we develop tools that allow us to enumerate the
different fixed points of a Boolean network, and to classify their
organisation. These tools are borrowed from statistical physics, based
on the identification of fixed points with the zero-energy ground
states of Hamiltonian (\ref{eq:H}). More precisely, we use the cavity
method \cite{MPZ, Libro_Martin} as recently developed in the framework
of disordered systems in order to solve statistical mechanics problems
defined on finite connectivity graphs, and reformulate it in a way able to
deal with Boolean Networks.

As seen in Sec.~\ref{sec:cc}, we are to identify an extended region in
the $(x, \alpha)$ plane, where an extensive {\em computational core}
of BN exists. Inside this region the mechanism of regulation -- given
not only by trivial cascades of logical implications, but also by non
trivial loop structures -- starts playing a relevant role. More
sophisticated techniques are needed to study this phase.

\subsection{Information flow in the network and definition of cavity biases}

The first step is to transform the Hamiltonian of Eq.~(\ref{eq:H})
from Boolean variables $x_i\in\{0,1\}$ to Ising spin $s_i=2x_i-1
\in\{\pm 1\}$. It is easy to see that each energy contribution in
Eq.~(\ref{eq:H}) can be rewritten as
\begin{eqnarray}
\label{eq:Handor}
E_{J_1,J_2, J_3}(s_1,s_2,s_3) &=& 
1-J_1 s_1 \left[ 1 - \frac{(1+J_2 s_2)(1+J_3 s_3)}2
\right]\,\,\,\,\,\,\,\,\,\,\,\,\mathrm {AND-OR\ class}\\
\label{eq:Hxor}
E_J(s_1,s_2,s_2) &=& 1-J s_1 s_2 s_3 \,\,\,\,\,\,\,\,\,\,\,\,\,\,\,
\,\,\,\,\,\,\,\,\,\,\,\,\,\,\,\,\,\,\,\,\,\,\,\,\,\,\,\,\,\,\,\,\,
\,\,\,\,\,\,\,\,\,\,\,\,\,\,\,\,\,\,\,\,\,\,\,\,\,\,\,\,\,
\mathrm{XOR\ class}\ ,
\end{eqnarray}
remember that the first entry ($s_1$) in these interaction terms
corresponds to the output of the Boolean function, whereas the others
($s_{2,3}$) describe the inputs.  

\begin{figure}[htb]
\vspace{0.2cm}
\begin{center}
\includegraphics[width=0.65\columnwidth]{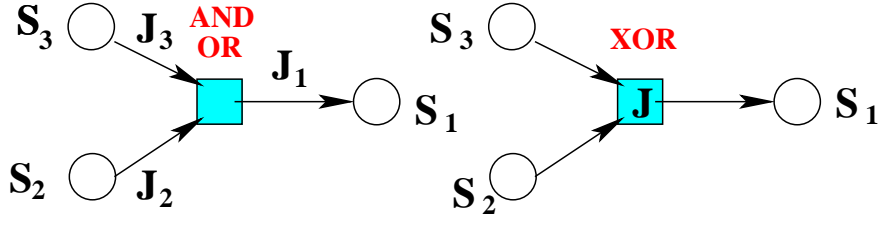}
\end{center}
\caption{AND-OR , XOR function nodes.}
\label{fig:function_node}
\end{figure}

In the first of these two equations, $J_1 = \pm 1$ switches between
the OR ($J_1=+1$) and the AND ($J_1=-1$) sub-families, while the other
parameters $J_2,J_3 = \pm 1$ act as negation of the input
variables. As an example, the OR function is given by
$(J_1,J_2,J_3)=(1,-1,-1)$. It is easy to check that the zero-energy
configurations are $(s_1,s_2,s_3)=(-1,-1,-1)$ (corresponding to two
false input variables and one false output variable), and $(1,-1,1)$,
$(1,1,-1)$, $(1,1,1)$ (corresponding to at least one true input
variable, and consequently a true output variable). The remaining four
configurations lead to energy two, and are thus forbidden in a
zero-energy ground state. There is only a unique parameter $J$ in
Eq.~(\ref{eq:Hxor}): it implements the XOR function ($J=1$) or its
negated counterpart $\overline{\mathrm{XOR}}$ ($J=-1$). The
parametrisation is displayed schematically in
Fig.~\ref{fig:function_node}.

In the cavity formalism developed originally in \cite{PM1,PM2}, one
sets up a self-consistent iterative procedure which allows to infer
the marginal probabilities for single spins or small groups of them
from the original BN. In each step, a cavity is carved into the
network by taking out one variable $s_i$, and all the Boolean
constraints $E_a$ containing it either as an input or output variable
(denoted formally as $a\in i$). The resulting reduced network is
called the cavity network BN$_i$. Further on, the properties of the
neighbours $s_j$ of $s_i$ in all ground-state configurations of BN$_i$
are characterised by integer {\it cavity fields} $h_{j\to a}$ (with
$a\in i,\ j\in a\setminus i$). More precisely, the minimum energy of
BN$_i$ with arbitrarily fixed $s_j$ can be written as
\begin{equation}
const - \sum_{j:\ j\in a\setminus i,\ a\in i} h_{j\to a} s_j \ .
\end{equation}
This equation contains the basic assumption of the cavity method: all
spins $s_j$ on the cavity system are statistically independent. In
large random networks this is expected to be asymptotically true due
to their locally tree-like structure, in finite graphs this may still
be a reasonable approximation. An immediate consequence of this
equation on all minimum-energy configurations of BN$_i$ including the
optimisation over all $s_j$ can be drawn: a positive cavity field
$h_{j\to a}>0$ indicates that the corresponding variable $s_j$ is
fixed to $+1$ on the cavity network BN$_i$, $h_{j\to a}=0$ indicates
that $s_j$ can take both values $\pm 1$, whereas $h_{j\to a}<0$
implies $s_j=-1$.

\begin{figure}[htb]
\vspace{0.2cm}
\begin{center}
\includegraphics[width=0.65\columnwidth]{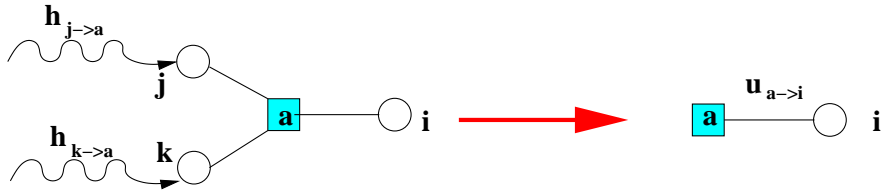}
\end{center}
\caption{The cavity bias $u$ sent from function node $a$ to site $i$
as a function of the cavity fields $h_{j \rightarrow a}$ and
$h_{k\rightarrow a}$.}
\label{fig:u_of_h}
\end{figure}

Imagine now, that we put back the spin $s_i$ and one constraint $E_a$
into the cavity graph, and that we fix $s_i$ to one of its two values.
The energy shift in ground state energy is (up to a constant) given
by, cf. Fig.~\ref{fig:u_of_h}:
\begin{eqnarray}
\label{minimisation}
\min_{s_j, s_k} E_{a}(s_i, s_j, s_k) -
(h_{j \rightarrow a }s_j +  h_{k \rightarrow a } s_k) &=& -w_{a
  \rightarrow i}^{\rm in} - s_i u_{a \rightarrow i}^{\rm in}  
\nonumber \\
\min_{s_i, s_j} E_{a}(s_i, s_j, s_k) -
(h_{i \rightarrow a } s_i +  h_{j \rightarrow a } s_j) &=& -w_{a
  \rightarrow k}^{\rm out} - s_k u_{a \rightarrow k}^{\rm out}  \nonumber \\
\min_{s_i, s_k} E_{a}(s_i, s_j, s_k) -
(h_{i \rightarrow a } s_i +  h_{k \rightarrow a } s_k) &=& -w_{a
  \rightarrow j}^{\rm out} - s_j u_{a \rightarrow j}^{\rm out}  
\end{eqnarray}

The $u$ are called {\it cavity biases}. Its physical meaning is clear:
in a minimum-energy configuration, the variable has to take a value
parallel to the bias.  Note that the initial BN is directed, as $s_i$
is given by the output value of a Boolean function $F_a$. Here we see
that cavity fields and biases are defined also in the directions
opposite to the directed nature of the BN. Within the notation of
Eq.~(\ref{minimisation}), in-biases act into the direction of the links
of the factor graph, out-biases against it. In analogy, we define
out-fields to act along the graph direction, in-fields against
\footnote{The apparent difference in the in-out notation between
cavity fields and biases is that the first are defined to act from a
variable to a constraint, the latter from a constraint to a
variable. The index ``in'' is, however, given to both message types on
factor graph links going from a Boolean function to its output, the
index ``out'' for links going from a variable to a function depending
on it.}.As a general consideration, it is not surprising that there
should be cases when information flows against the directed networks
arrows: imagine a configuration in which $s_1 = 1$ being output of an
AND gate. This will immediately imply that $s_2 = s_3 = 1$, since this
is the only input configuration leading to the desired
output. Back-propagation of information is therefore plausible. We will
see that there are cases where both conditions (presence or
absence of back propagation of information) are met.
%Minimizing the
%energy shifts should ensure us that the messages (fields and biases)
%we are calculating are the ones that make variables to point in the
%direction that maximize the number of satisfied constraits.
  
At this point we can explicitly perform the minimisation in the above
equations and express the $w$s and $u$s as functions of the messages
$h_{j \rightarrow a }$. Using the notation of Eqs.~(\ref{eq:Handor})
and (\ref{eq:Hxor}), we obtain for the two functional classes:
\begin{itemize}
\item Canalising Functions (AND-OR family):
\begin{eqnarray}
\hat u_{\vec J}^{\rm out}(h_{2 \rightarrow a},h_{3 \rightarrow a}) &=& 
J_1\left( \theta(-J_3 J_2 h_{2 \rightarrow a } h_{3 \rightarrow a}) - 
\theta(J_3 h_{3 \rightarrow a }) \right) \nonumber \\
w_{\vec J}^{\rm out}(h_{2 \rightarrow a },h_{3 \rightarrow a }) &=&
|h_{2\rightarrow a }| + |h_{3\rightarrow a }| -
|\hat u^{\rm out}_{\vec J}(h_{2 \rightarrow a }, h_{3 \rightarrow a })|  - 2  
\theta(-J_2 J_3 h_{2\rightarrow a } h_{3 \rightarrow a }) \theta(J_3
h_{3 \rightarrow a }) 
\nonumber \\
\hat u_{\vec J}^{\rm in}(h_{2 \rightarrow a },h_{3 \rightarrow a }) &=& 
J_1( \theta(-J_2 h_{2 \rightarrow a }) + \theta(-J_3 h_{3 \rightarrow a }) 
\nonumber - \theta(J_2 J_3 h_{2 \rightarrow a } h_{3 \rightarrow a } ) )
\nonumber \\
w_{\vec J}^{\rm in}(h_{2 \rightarrow a },h_{3 \rightarrow a }) &=& |h_{2
  \rightarrow a }| + |h_{3 \rightarrow a }| -
|\hat u^{\rm in}_{\vec J}(h_{2 \rightarrow a }, h_{3 \rightarrow a })| 
\label{bias:AND}
\end{eqnarray}
\item Non-Canalising Functions (XOR family):
\begin{eqnarray}
\hat u_{J}^{\rm out}(h_{2 \rightarrow a},h_{3 \rightarrow a})
 &=& J Sign(h_{2\to a} h_{3 \rightarrow a})\nonumber\\
 w_{J}^{\rm out}(h_{2 \rightarrow a},h_{3 \rightarrow a}) &=&
 |h_{2 \rightarrow a}| + |h_{3 \rightarrow a}| - 
 |\hat u^{\rm out}_{J}(h_{2 \rightarrow a}, h_{3 \rightarrow a})| \nonumber \\
 \hat u_{J}^{\rm in}(h_{2 \rightarrow a},h_{3\rightarrow a}) &=& J Sign(h_{2
 \rightarrow a} h_{3 \rightarrow a})\nonumber \\
 w_{J}^{\rm in}(h_{2\rightarrow a},h_{3\rightarrow a}) &=& |h_{2\rightarrow
 a}| + |h_{3 \rightarrow a}| - 
|\hat u^{\rm in}_{J}(h_{2 \rightarrow a}, h_{3 \rightarrow a})|
\label{bias:XOR}
\end{eqnarray}
\end{itemize}
Since flipping the spin on the right-hand side of
Eqs.~(\ref{minimisation}) may lead to energy differences $0,\pm 2$,
cavity biases take integer values $\hat u^{\rm in,out} \in \{ -1, 0,
+1 \}$.

\label{sec:wp}
\begin{figure}[htb]
\vspace{0.2cm}
\begin{center}
\includegraphics[width=0.65\columnwidth]{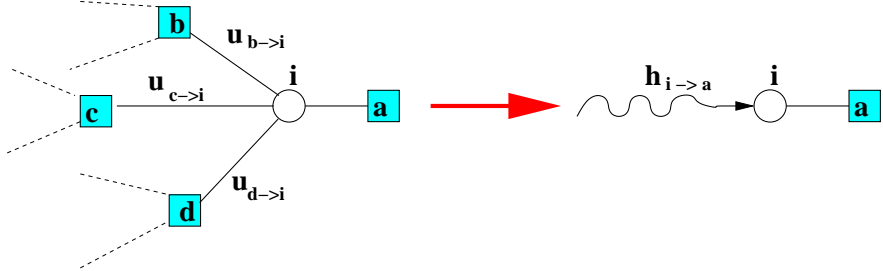}
\end{center}
\caption{The cavity field $h_{i \rightarrow a}$ is the sum of all
cavity bias pointing to site $i$ except that of function node $a$.}
\label{fig:h_eq_sum_u}
\end{figure}

In order to close the equations for the cavity fields and biases
self-consistently, we also have to give an expression of the $h$ in
terms of the $u$. The cavity field $h_{i\to a}$ can be easily determined
by putting back into the cavity graph BN$_i$ all but one constraints
($b\in i\setminus a$). The interpretation of the fields in terms of the
energy difference due to flipping $s_i$ leads immediately to
\begin{equation} 
h_{i\rightarrow a} = 
\sum_{b\in i \setminus a} u_{b\rightarrow i} \ ,
\label{eq:somma}
\end{equation}
cf.~\cite{PM1, PM2, MPZ,Libro_Martin}) and see
Fig.~\ref{fig:h_eq_sum_u}. The true field acting on $s_i$ in the full
problem can also be calculated from the cavity biases by putting back
all constraints:
\begin{equation} 
H_{i} = 
\sum_{a\in i} u_{a\rightarrow i} \ .
\label{eq:somma_vera}
\end{equation}
Eqs. (\ref{bias:AND}-\ref{eq:somma_vera}) are called {\it warning
propagation equations} (WP). They can be exploited algorithmically:
first Eqs. (\ref{bias:AND}-\ref{eq:somma}) are iterated until a
solution for all cavity fields and biases is found, and then the $H_i$
are determined. As discussed above, a positive field forces the
corresponding spin to take value $+1$, a negative value $-1$. Only
sites with zero fields are not yet fixed in value, and can be fixed
via a graph decimation process, cf.~\cite{MPZ,Libro_Martin}.

This iterative solution works, however, only if there are at most a
few solutions to the WP equations, corresponding in a physical
language to only one or a few thermodynamic states as zero
temperature. This assumption is equivalent to replica symmetry.  We
will discuss this case in detail before coming to the possibility of
multiple solutions to WP and broken replica symmetry.

\subsection{Replica symmetric solution (warning propagation)}
    
One can push forward the analysis considering the average over the
random BN ensemble specified by the average density $\alpha$ of
functions per variable, and the fraction $x$ of XOR-type Boolean
functions. In this case we can define two probability distributions
$Q^{\rm out}(u)$ and a $Q^{\rm in}(u)$ representing the average
probability of finding an out/in bias of strength $u$ and satisfying
the following self-consistent set of integral equations:
\begin{eqnarray}
\label{eq_self_consistency_0}
 Q^{\rm out}(u) &=& \left \langle \int dh dg P^{\rm out}_{k,b}(h) 
P^{\rm in}_{k'}(g)\,\,
\delta\left( u -
 \hat u_{\{J\}}^{\rm out}(h,g)\right ) \right \rangle_{k,k',b,\{J\}}
\nonumber \\
Q^{\rm in}(u) &=& \left \langle 
\int dh dg P^{\rm out}_{k,b}(h) P^{\rm out}_{k',b'}(g) 
\,\,\delta\left(u - \hat u_{\{J\}}^{\rm in}(h,g)\right)
 \right \rangle_{k,k',b,b',\{J\}}
\nonumber \\
\end{eqnarray}
defining 
\begin{eqnarray}
\label{eq_self_consistency}
P^{\rm in}_{k}(h) &=& \int \prod_{a=1}^{k} du_{a} Q^{\rm out}(u_{a})\,\,
\delta\left(h - \sum_{a=1}^k u_{a}\right) 
\nonumber \\
P^{\rm out}_{k,b}(h) &=& \int du Q^{\rm in}(u) \prod_{a=1}^{k} 
du_{a} Q^{\rm out}(u_{a})\,\,
\delta\left(h - \sum_{a=1}^k u_{a} -b u \right)\,\,\,\,\, 
\end{eqnarray}
and where $\langle \cdot \rangle_{k,b,\{J\}}$ is the quenched average over
$\rho^{\rm out}( k ) = \exp(-K\alpha ) (K \alpha)^k/k!$ (the site
out-degree distribution), $\rho^{\rm in}(b) = \alpha \delta(b;1) +
(1-\alpha)\delta (b;0)$ (the site in-degree distribution) and the
probability distribution over the different function node types
$p(\{J\})$. 

Assuming that the $p(\{J\})$ is flat for both canalising
and non canalising function, we force fields the distribution to be symmetric
under $u \leftrightarrow -u$. Recalling that $u \in \{ -1, 0, 1\}$
one can assume that a single number $\eta\in(0,1)$ can parametrise the
cavity bias distribution:
\begin{eqnarray}
\label{eq:q_of_u}
Q^{\rm out}(u) &=& \eta^{\rm out} \delta(u) + \frac{(1-\eta^{\rm out})}{2} 
[\delta(u-1) +\delta(u+1)] \nonumber\\
Q^{\rm in}(u) &=& \eta^{\rm in} \delta(u) + \frac{(1-\eta^{\rm in})}{2} 
[\delta(u-1) +\delta(u+1)] 
\end{eqnarray}
Inserting this simple functional form in
Eqs.~(\ref{eq_self_consistency}), and after some algebra, one obtains
the following relations, for the graph averaged $P^{\rm in,out}(h) =
\langle P^{\rm in,out}_{k,b}(h) \rangle_{k,b}$:
\begin{eqnarray}
\label{eq:p_of_h}
P^{\rm in}(h) &=& e^{-K\alpha(1-\eta^{\rm out})} 
I_h\left(K\alpha (1-\eta^{\rm out})\right)
\nonumber\\
P^{\rm out}(h) &=& e^{-K\alpha(1-\eta^{\rm out})} \left\{ \eta^{\rm in}
I_h\left( K\alpha (1-\eta^{\rm out})\right) + (1-\eta^{\rm in})
\left[ I_{h+1}(K\alpha (1-\eta^{\rm out})) + 
I_{h-1}(K\alpha (1-\eta^{\rm out}))\right]
\right \}
\end{eqnarray}
where $I_h(x) := \sum_{l=h}^\infty (x/2)^{2l}/[(l-h)!l!]$ is the
$h^{th}$-order modified Bessel function. One can close analytically
the system of equations considering the weight in zero of the two
$P(h)$ in the two limiting cases:
\begin{itemize}
\item A graph made of only canalising functions ($x=0$):
\begin{eqnarray}
\label{eq_rs_andor}
2\eta^{\rm out} -1 &=& \sqrt{1-\eta^{\rm in}} 
\left[ 1 - e^{-2\alpha (1-\eta^{\rm out})} \right]
\nonumber \\
\eta^{\rm in} &=& e^{-2\alpha(1-\eta^{\rm out})}
\left\{ I_0(2\alpha(1-\eta^{\rm out})) 
-\alpha(1-\eta^{\rm in})\left[I_{0}(2\alpha(1-\eta^{\rm out}))-I_{1}
(2\alpha(1-\eta^{\rm out}))\right]\right\}
\end{eqnarray}

\item A graph made of only non-canalising functions ($x=1$):
\begin{eqnarray}
\label{eq_rs_xor}
1-\eta^{\rm out} &=& \sqrt{1-\eta^{\rm in}} \left[ 
1 - e^{-2\alpha (1-\eta^{\rm out})}\right]
\nonumber \\
1- \sqrt{1-\eta^{\rm in}} &=& e^{-2\alpha(1-\eta^{\rm out})}
\left\{ I_{0}(2\alpha(1-\eta^{\rm out})) 
-\alpha(1-\eta^{\rm in})\left[I_0(2\alpha(1-\eta^{\rm out}))-I_{1}
(2\alpha(1-\eta^{\rm out}))\right]\right\}  
\end{eqnarray}
\end{itemize}
These equations are known as {\em density evolution equations} in
Information Theory \cite{RichUrbanke01}.  However, it turns out that
the only solution for both equations is the trivial paramagnetic one,
{\it i.e.}  $\eta^{\rm in} = \eta^{\rm out} = 1$. Therefore, WP gives
no interesting information in order to find fixed points (there are no
biases), nor on their number, nor on their separation in phase space.

\subsection{Belief Propagation}
\label{sec:bp}
To overcome this limitation, we have to go beyond the WP approximation
introduced so far. The idea is to introduce in the study of BN real
valued messages in $[0,1]$ measuring the {\it probability} that a
variable takes a certain value in the ground states of the (cavity)
problem. This resolves finer the case of vanishing cavity fields and
biases - according to the above solution this means of all of them.
We can introduce a refined iterative method -- called Belief
Propagation (BP) \cite{Yedidia,SumProd,BMZ} -- that allows to define
the fraction of zero energy configurations having, say, $x_i = 0$. It
is a rather general algorithm that allows to calculate marginal
probabilities of problem defined on factor graphs. Note that also BP
is based on the assumption of replica symmetry, {\it i.e.} of the
existence of a single pure state.

Calling $a$ one of the clauses where $x_i$ appears and
$E_a(\{x_l\}_{l\in a}) = 2 x_a \oplus F_a(x_{a_1,\dots, a_K}) $ as in
Eq.~(\ref{eq:H}), we can introduce the quantities:

\begin{itemize}
\item $\mu_{i\rightarrow a}(x_i)$: the probability that variable $i$ takes 
value $x_i$ in the absence of clause $a$.
\item $m_{a \rightarrow i}(x_i)$: the non-normalised probability that
clause $a$ is satisfied when variable $i$ takes value $x_i$.
\end{itemize}

The two quantities satisfy the following set of equations:
\begin{eqnarray}
\label{eq:bp}
m_{a\rightarrow i}(x_i)  &=& \sum_{\{x_l\}_{l \in a \setminus i}} 
\left[1- \frac 12 E_a (\{x_l\}_{l\in a})\right] 
\prod_{l \in a \setminus i} \mu_{l\rightarrow a} (x_l) 
\nonumber\\
\mu_{i\rightarrow a}(x_i) &=& C_{i\rightarrow a} \prod _{b\in i
  \setminus a } m_{b\rightarrow i}(x_i)
\end{eqnarray}
where $C_{i\rightarrow a}$ is a constant enforcing the normalisation
of the probability distribution $\mu_{i\rightarrow a}(x_i)$.  The
iteration of Eq.~(\ref{eq:bp}) provides the correct marginal on a
tree, where it can be shown that the full probability measure ${\cal
P}(\vec x)$ of configurations $\vec x\in \{0,1\}^N$ satisfying all
constraints can be expressed as a product of site and clause marginals
\cite{SumProd,BMZ}:
\begin{equation}
\label{eq:full_prob}
{\cal P}(\vec x) = \prod_{a \in A } P_{a} ( \{x_l\}_{l\in a} ) \prod_{i=1}^N P_i
( x_ i ) ^ {1-d_i}
\end{equation}
where the marginal probability distribution $P_i(x_i)$ of variable $i$
is the fraction of zero-energy configurations having variable $i$ set
to value $x_i$, and the joint probability distributions $P_a(\{
x_l\}_{l\in a} )$ of all variables belonging to function node $a$ is
the fraction of zero-energy configurations having variables ${i\in a}$
set to $\{x_i\}_{i\in a}$. Further on $d_i$ denotes the total degree
of variable $i$, summing its in- and out-degree.

The Entropy $S$, {\em i.e.}  the logarithm of the number of fixed
points ${\cal N}_{\mathrm{fp}}$, can be also computed in terms of the
marginal introduced in Eq.~(\ref{eq:full_prob}) as:
\begin{equation}
S = - \sum_{\vec x}{\cal P}(\vec x)\ln {\cal P}(\vec x) = -\sum_{a \in A\ } 
\sum_{\ \{x_l\}_{l \in a}} P_{a} ( \{x_l\}_{l\in a}) 
\ln  P_{a} ( \{x_l\}_{l\in a} ) +\sum_{i=1}^N \sum_{x_i}  ( d_i-1 ) P_i
( x_ i ) \ln  P_i (x_i)
\end{equation}
We can now express the marginals in terms of the messages introduced
in Eq.~(\ref{eq:bp})
\begin{eqnarray}
\label{eq:marginals}
P_{a}(  \{x_i\}_{i\in a} )  &=& c_{a}\left[1-\frac12 
E_a(\{x_i\}_{i\in a})\right] \prod_{i \in a } 
\mu_{i\rightarrow a}(x_i)\nonumber \\
P_i ( x_i )  &=& c_i \prod_{a\in i}  m_{a\rightarrow i}(x_i)
\end{eqnarray}

It is convenient to parametrise probabilities $\mu_{i\rightarrow
a}(x_i)$ in terms of numbers $\eta_{i\rightarrow a}\in[0,1]$ as
$\mu_{i\rightarrow a}(x_i) = \eta_{i\rightarrow a} \delta ( x;0 ) + (
1-\eta_{i\rightarrow a}) \delta (x;1)$ and then eventually express
the entropy $S$ in terms of message quantities:
\begin{equation}
S = -\sum_{a\in A} \ln c_a + \sum_{i\in X} (d_i-1) \ln c_i -
\sum_{a\in A} \sum_{i\in a} C_{i\rightarrow a}  
\end{equation}

In analogy with the case of Warning Propagation, it is convenient to
classify the set of messages $\{ m \}$ (one for each edge) as type in
and type out, according to their orientation with respect to the
graph.  WP equations can be retrieved as a particular case where on
each link either $\eta_{i\rightarrow a} \in \{0,1\}$ or
$\eta_{i\rightarrow a} = 1/2$. In particular, we have seen in the
previous paragraph how the only WP solutions are the completely
unbiased ones, corresponding to $\eta_{i\rightarrow a} = 1/2$ on every
link.

Let $\{m^{\rm out}\}\subset \{m\}$ be the subset messages flowing
against the orientation of the graph, {\em i.e.}  messages flowing
from functional node $a$ to the regulatory sites $s_2$ and $s_3$, as
the case displayed in Fig.~\ref{fig:function_node}. Let also $\{m^{\rm
in}\}\subset \{m\}$ be the subset of messages flowing from functional
nodes $a$ to the regulated site: as in the case of $m_{a \rightarrow
1}$ as displayed in Fig.~\ref{fig:function_node}. Obviously $\{m\} =
\{m^{\rm out} \} \cup \{m^{\rm in} \}$.

The directed nature of the graph shows unexpected statistical
properties of the solutions of the BP Eqs.~(\ref{eq:bp}) in all cases
where zero energy ground states exist. It turns out that the set of
fixed points obtained by iterating until convergence
Eqs.~(\ref{eq:bp}), is characterised by the property that for each
link $a\rightarrow i$, $m_{a\rightarrow i}^{\rm out} ( 0 ) =
m_{a\rightarrow i}^{\rm out} ( 1 )$, {\em i.e.}  the set $\{ m^{\rm
out} \}$ is again completely unbiased \footnote{It is easy to check
that such a solution is consistent with the BP equations. It is,
however, non-trivial that it is the only one.}. The only biased
messages belong to $\{ m^{\rm in} \}$. Thus, BP cannot see back
propagation of information on this type of directed networks.  In this
case an interesting series of simplifications hold:
\begin{itemize}
\item Since messages $\{m\}$ are non normalised, we can set both
  components of messages belonging to  $\{ m^{\rm out} \}$ to 1.
\item The marginal $P_i(x_i)$ is then specified by the only
  contribution of message in: $P_i(x_i) = c_i m^{\rm in}_{a \rightarrow
  i} (x_i) $, where $a$ is the function node regulating $i$.
\item Eqs.~(\ref{eq:bp}) can be therefore expressed in a more compact
form:
\begin{equation}
\label{eq:bpdir}
P_i(x_i ) = \sum_{\{ x_l\}_{l\in a\setminus i}\ } \prod_{\ l\in a\setminus i} P_l (
x_l) \left[1-\frac 12 E_a(x_i, \{x_l\})\right]\,\,\, .
\end{equation}
\end{itemize}
It is interesting to note that in the present case of directed graphs,
one can directly iterate marginal probabilities instead of passing
through the intermediate step of cavity messages $\{m,\mu\}$ as
happens to BP equations on general graphs. The rationale for this
simplification is inherent to the peculiar asymmetric form of $E_a = 2
x_a \oplus F_a(x_{a_1},...,x_{a_K})$ that makes the factorisation of
the marginal
\begin {equation}
\label{eq:factdir}
P(x_a, x_{a_1},...,x_{a_K}) = P(x_a | x_{a_1},...,x_{a_K}) P(
x_{a_1},...,x_{a_K} ) = \left[1- \frac 12 E_a ( x_a,
x_{a_1},...,x_{a_K} )\right] \prod_{l=1}^K P_{a_l}(x_{a_l})
\end{equation}
exact on a tree (again iff only zero energy configurations are taken
into account, otherwise the last step in Eq.~(\ref{eq:factdir}) would
not hold). By inserting the factorised form (\ref{eq:factdir}) into the
the full probability distribution ${\cal P}(\vec x)$ of
Eq.~\ref{eq:full_prob}, one can show, after some simple algebra, that:
\begin{equation}
{\cal P}(\vec x) = \prod_{a \in A} \left[1-\frac 12 
E_a ( x_a,x_{a_1},...,x_{a_K} )\right]
\prod_{i \in \mathrm {EIV}} P_i( x_i )  
\end{equation}
where EIV is the set of {\em external input variables}, {\em i.e.} the set
of all $N-M$ non-regulated variables (see Fig.~(\ref{fig:sandwich})). In
the case that non-regulated variables are not biased by external
fields, we have $P_i(0) = P_i(1) = 1/2$ for all $i\in \mathrm{EIV}$. The
entropy then turns out to be, with some simple algebra:
\begin{eqnarray}
\label{eq:entropybp}
S &=& - \sum_{\vec x}\left\{ \prod_{a \in A} 
\left[1-\frac 12 E_a ( x_a,x_{a_1},...,x_{a_K} )\right]
\prod_{i \in \mathrm {EIV}} P_i( x_i ) 
\left [ \sum_{a\in A} \ln\left[1-\frac 12 E_a
  ( x_a,x_{a_1},...,x_{a_K} )\right] + \sum_{i\in \mathrm{EIV}}
  \ln P_{i}(x_i)  \right] \right\} \nonumber \\
&=& (N-M)\ln (2) 
\end{eqnarray}

\begin{figure}[htb]
\vspace{0.2cm}
\begin{center}
\includegraphics[width=0.65\columnwidth]{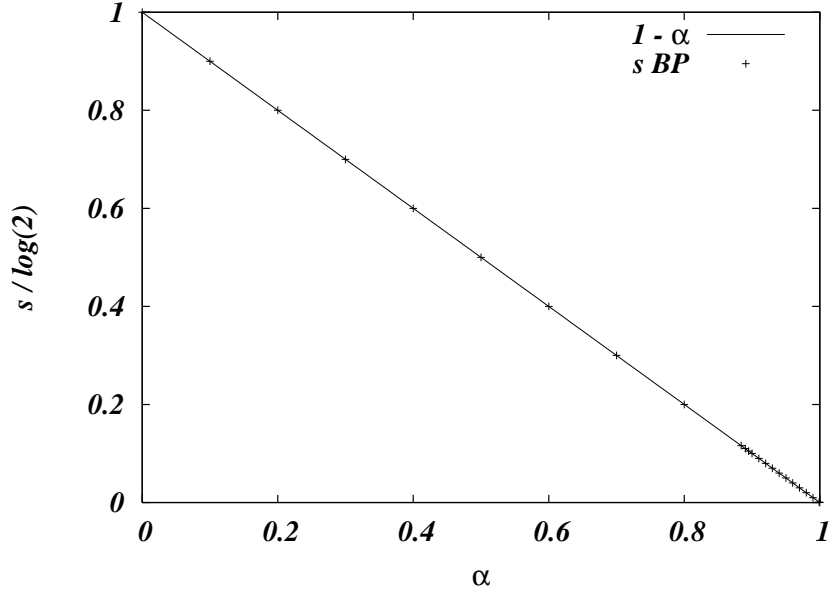}
\end{center}
\caption{Entropy density $s = S/N$ vs. average graph connectivity
  $\alpha$. Dots correspond to the average value obtained by running
  the BP equations on samples of $N=100 000$ at different node
  concentrations $x$. It turns out that the entropy $s$ is not depending
  on how one fix the different function node type but only on the
  topology of the graph, {\em i.e.} just on the density  of non
  regulated sites $(N-M)/N = 1-\alpha$.}
\label{fig:entropybp}
\end{figure}

This result implies that fixed points satisfying all Boolean functions
exist for all $N\geq M$ {\rm i.e.} for $\alpha<1$. The line $\alpha=1$
can be then considered as the SAT/UNSAT transition line. The entropy
value shows that each configuration of the $(1-\alpha)N$ external
input variables leads, on average, to one stationary point, {\em i.e.}
the state of all internal and functional variables is mainly
determined by the external condition, even in the case of a PER core
(see Sec.~\ref{sec:PER}), where this fixing is not just a simple
logical implication. This result is far from being rigorous, since we
do not know at present how to prove that the only unbiased messages
are those flowing from functional nodes to regulated sites. We have
checked numerically the result presented in Eq.~(\ref{eq:entropybp})
running the BP equations on single sample at different values of
$\alpha$ as displayed in Fig.~\ref{fig:entropybp}.

\subsection{Survey Propagation}

Previous approaches rely on the assumptions that all zero energy
configurations are arranged into one single cluster in phase space,
{\it i.e.}~that cavity fields and biases are well-defined on the basis
of the WP equations alone. However, if the phase space breaks into
many pure states ({\em i.e.} macroscopically separated clusters of
zero-energy configurations), each of these states corresponds to one
non-trivial solution of WP. With high probability, a simple iteration
of the WP equations does not converge to one of these solutions, but
just to the trivial one. The relevant order parameter of the model is
therefore not given by a single field/bias per link, but by the
distribution of both fields and biases. This distribution is not to be
confused with the average order parameter described by the density
evolution equations, but it is a broad distribution over the values of
the biases (fields) sampled over all clusters (pure states) for one
network realisation. In the simple case of a single pure state, {\it
i.e.}~in the replica symmetric phase, single-site probability
distributions become delta functions and the order parameter
simplifies to a single global probability distribution. In the more
general case, the order parameter averaging over the graphs ensemble
becomes a probability distribution of probability distributions. Let
us start defining (with an abuse of notation with respect to replica
symmetric equations) the {\em survey} $Q_{a\rightarrow i}(u)$ from
function node $a$ to site $i$ as the histogram of cavity bias
$u_{a\rightarrow i}$ over all ${\cal N}_{cl}$ different clusters:
\begin{equation}
Q_{a\rightarrow i}(u) = \frac{1}{{\cal N}_{cl}} \sum_{l=1}^{{\cal N}_{cl}}
\delta ( u^l_{a\rightarrow i } - u ) 
\end{equation}
The $Q$ distributions are self-consistently computed via the Survey
Propagation equations (SP) \cite{MPZ,Libro_Martin,BMZ}:
\begin{eqnarray}
\label{eq:sp}
P_{i\rightarrow a}(h) &=& C_{i\rightarrow a} \int 
\left[ \prod_{b\in i \setminus a}
du_{b\rightarrow i}\  Q_{b\rightarrow i}(u_{b\rightarrow i})
  \right]
\delta \left( h - \sum_{b\in i \setminus a} u_{b\rightarrow i} \right) 
e^{-y \Delta E^{iter}_{i\rightarrow a}( \{u_{b\rightarrow i}\} ) }
\nonumber \\
Q_{a\rightarrow i}(u) &=& \int \left[ 
\prod_{l\in a \setminus i } dh_{l\rightarrow a}\
P_{l\rightarrow a}(h_{l\rightarrow a}) \right]
\delta \left( u - \hat u_{a\rightarrow i} 
(\{h_{l\rightarrow a} \}_{l \in a \setminus i}) \right) 
\end{eqnarray}
where $C_{i\rightarrow a}$ are normalisation constants, and both
functions $\hat u$ are defined in Sec.~\ref{sec:wp}. 
\begin{equation}
\label{eq:shift}
\Delta E^{iter}_{i\rightarrow a}( \{u_{b\rightarrow i}\} ) = 
\sum_{  b\in i\setminus a} \left\vert u_{b\rightarrow i} \right\vert - 
\left\vert\sum_{  b\in i\setminus a} u_{b\rightarrow i} \right\vert
\end{equation}
represents the energy shift due to re-insertion of site $i$ and
constraint $a$ into the cavity graph, along the iteration/minimisation
procedure described in Eqs.~(\ref{minimisation}), (\ref{bias:AND}) and
(\ref{bias:XOR}). The appearance of the reweighting parameter $y>0$
allows therefore to scan different energy levels of metastable
states. It acts similarly to the inverse temperature in the usual
Boltzmann weight: upon increasing the value of $y$, the probability
measure concentrates on decreasing levels of energy, and the limit
$y\rightarrow \infty$ corresponds to zero-energy ground states. Indeed,
one can relate $y$ to the derivative of the {\em complexity} $\Sigma$
with respect to the energy shift, where the complexity is defined in
analogy to the entropy as the logarithm of the number of thermodynamic
states at given energy. In this case, forcing $y$ to very large values
implies that the only configurations that contributes to surveys are
those with vanishing energy shifts.

The above formalism is formulated for a single sample analysis, and in
this form it allows for a straightforward implementation of the SP
algorithm, which amounts to iterate Eq.~(\ref{eq:sp}) on real
networks, for each function node $a\in A$ and each variable $i$
involved in $a$, from a random initial condition until
convergence. However, it is easy to generalise the SP equations to
deal with average quantities on the random graph ensemble introduced
in Sec.~\ref{sec:model}. General considerations on the existence of a
well defined thermodynamic limit \cite{GuerTon,FranLeo} imply the
existence of a functional probability measure ${\cal Q}[Q(u)]$
describing the statistics of the $Q(u)$ on the different
clusters. Again, as in the case of the WP and BP equations of
Sec.~\ref{sec:wp} and \ref{sec:bp} respectively, it turns out to be
convenient to classify the surveys as type in and out. The
self-consistency equations for the ${\cal Q}[Q^{\rm out}(u)]$ and
${\cal Q}[Q^{\rm in}(u)]$ then become:

\begin{eqnarray}
\label{eq:1rsbave}
{\cal Q}[Q^{\rm out}(u)] &=& 
\left\langle
\int d {\cal Q}[Q^{\rm in}]\prod_{a=1}^{k} d {\cal Q}[Q^{\rm out}_a]
\prod_{a'=1}^{k'} d{\cal Q}[\tilde Q^{\rm out}_{a'}]
\,\,\, \hat \delta 
\left[ Q^{\rm out}(u) - 
Q^{\rm out}\left(u\,\,\big\vert\,\, \{Q^{\rm out}_a\},\{\tilde 
Q^{\rm out}_{a'}\}, Q^{\rm in}\right) 
\right]
\right\rangle_{k,k'} \\
{\cal Q}[Q^{\rm in}(u)] &=& 
\left\langle
\int d {\cal Q}[Q^{\rm in}] d{\cal Q}[{\tilde Q}^{\rm in}]
\prod_{a=1}^{k} d{\cal Q}[Q^{\rm out}_a]
\prod_{a'=1}^{k} d{\cal Q}[\tilde Q^{\rm out}_{a'}]
\,\,\, \hat \delta 
\left[ Q^{\rm in}(u) -  Q^{\rm in}\left(u\,\,\big\vert\,\, 
\{Q^{\rm out}_a\},\{\tilde Q^{\rm out}_{a'}\}, 
Q^{\rm in}, \tilde Q^{\rm in} \right) 
\right]
\right\rangle_{k,k'}  \nonumber
\end{eqnarray}
where $d{\cal Q}[Q^{i,o}] := {\cal Q}[Q^{i,o}] dQ^{i,o} $ denotes the
probability distribution of the probability $Q^{i,o}$, {\em i.e.}~the
integrals in Eq.~(\ref{eq:1rsbave}) are done over all probability
distributions $Q^{i,o}$ with weight ${\cal Q}[Q^{i,o}]$), and $\hat
\delta[.]$ is a functional delta.
\begin{eqnarray}
\label{eq_self_consistency-1rsb}
&&Q^{\rm out}\left( u \,\, \big \vert \,\, 
\{Q^{\rm out}_a\},\{\tilde Q^{\rm out}_{a'}\}, Q^{\rm in} \right)  =  
\left\langle \int dh dg P_{k,b}^{\rm out}(h) P^{\rm in}_{k'}(g) 
\,\,\,\delta\left( u - \hat u_{\{J\}}^{\rm out}(h,g)\right)\right\rangle_{b,\{J\}}
\nonumber \\
&&Q^{\rm in}\left(
u \,\, \big \vert  \,\,
\{Q^{\rm out}_a\},\{\tilde Q^{\rm out}_{a'}\}, Q^{\rm in},
{\tilde Q}^{\rm in} \right) = 
\left\langle \int dh dg P^{\rm out}_{k', b'}(h) P^{\rm out}_{k,b}(g) 
\,\,\, \delta\left(u - \hat u_{\{J\}}^{\rm in}(h,g)\right) 
\right\rangle_{b,b', \{J\}}
\\
&&P^{\rm in}_{k}(h) = \frac{1}{B^{\rm in}_k} \int \prod_{a=1}^{k} du_{a} 
Q^{\rm out}_{a}(u_{a})
\delta\left (h - \sum u_{a}\right ) 
\exp\left\{-y\left(\sum_{a=1}^k |u_{a}| - \left\vert 
\sum_{a=1}^k u_{a} 
\right\vert 
\right)\right\}
\nonumber \\
&&P^{\rm out}_{k,b}(h) =  \frac{1}{B^{\rm out}_k} 
\int du Q^{\rm in}(u) \prod_{a=1}^{k} du_{a} Q^{\rm out}_{a}(u_{a})  
\delta\left(h - \sum_{a=1}^k u_{a} - b u \right) 
\exp\left\{-y\left(\sum_{a=1}^k |u_{a}| + |bu| - \left\vert
\sum_{a=1}^k u_{a} + b u\right\vert 
\right)\right\} \nonumber 
\end{eqnarray}
The parameter $y$ is the reweighting coefficient which takes into
account level crossing of states under cavity iterations, and
consequently we have added the normalisation factors
$B^{i,o}_k$. Since we are interested in zero energy configurations we
will consider the $y\rightarrow \infty$ limit, where the reweighting
factor filters out all non-SAT configurations.

Let us analyse first the pure XOR ($x=1$) case where the computation
can be be done analytically along the lines of \cite{XOR}. In the mixed
case, we were not able to write analytic formulae for the complexity,
hence we will present the single sample analysis. In the case $x=1$,
the numerical solution of Eq. (\ref{eq_self_consistency-1rsb})
indicates that there is a non trivial solution for large values of the
reweighting $y$ only in the region $\alpha > 0.883867$.

Numerically, in the $y \to \infty$ limiting case, one observes that
probability distributions of biases peak into functional forms
(\ref{eq:q_of_u}), with a fraction of trivial probability
distributions peaked on $\delta (u)$ and a fraction of non trivial
ones.  Therefore the order parameters can be parametrised via two
scalar probability distributions of the $\eta$'s variables
$\rho_{\rm out}(\eta)$ and $\rho_{\rm in}(\eta)$ that take the form:
\begin{eqnarray}
\label{distribution_eta}
\rho_{\rm out}(\eta) &=& r_{\rm out} \delta(\eta -1) 
+ (1-r_{\rm out}) \tilde \rho_{\rm out}(\eta)
\nonumber \\
\rho_{\rm in}(\eta) &=& r_{\rm in} \delta(\eta -1) 
+ (1-r_{\rm in}) \tilde\rho_{\rm in}(\eta)
\end{eqnarray}
where the $r$'s are the fraction of trivial cavity biases.  The non
trivial cavity biases are characterised by a distribution
$\tilde{\rho}$ which in the limit $ y \rightarrow\infty $ converges to
the delta function in $\eta=0$.

Looking at the self consistency equations
(\ref{eq_self_consistency-1rsb}), the only way one can obtain a non
trivial distribution $Q^{\rm in}(u)$ is when both $P^{\rm
out}_{k,b}(h) \neq \delta(h)$ and $P^{\rm out}_{k',b'}(g) \neq
\delta(g)$. The probability that the field $P^{\rm out}(0) = P^{\rm
in}(0)= 0$ is given by the probability of picking all $k$ neighbouring
$Q(u)=\delta(u) $ , {\em i.e.}  respectively $r_{\rm out}^k$ for
$P^{\rm in}(0)$ and $r_{\rm out}^k r_{\rm in}^b$ for $P^{\rm
out}(0)$. Putting everything together and averaging over the
Poissonian/Bimodal degree distribution defined in
Eqs.~(\ref{eq:degdistr}), we obtain:
\begin{eqnarray}
\label{pr_h_out_zero}
P^{\rm out}(0) &=& \left(e^{-2\alpha} \sum_{k=0}^\infty
\frac{(2\alpha)^k}{k!}r_{\rm out}^k\right) \left( \sum_{b=0}^1[\alpha
\delta(b;1) + (1- \alpha) \delta(b;0) ]r_{\rm in}^b \right) 
= e^{-2\alpha(1-r_{\rm out})}
\left(\alpha r_{\rm in} + 1-\alpha \right )\nonumber \\
P^{\rm in}(0)  &=& e^{-2\alpha} \sum_{k=0}^\infty
\frac{(2\alpha)^k}{k!}r_{\rm out}^k = e^{-2\alpha(1-r_{\rm out})} 
\end{eqnarray}  
Now we can close the equations on the $r$s, considering that the
probability of having a non-trivial $Q(u)$ is given by the probability
that none of the other two incoming distribution is trivial:
\begin{eqnarray}
\label{eq:transition_alpha}
1-r_{\rm out} &=& \left[1-P^{\rm out}(0)\right]
\left[ 1-P^{\rm in}(0) \right] = 
\left[ 1 - e^{-2\alpha(1-r_{\rm out})} \right] 
\left[ 1 -\left[1-\alpha(1-r_{\rm in})\right]  
e^{-2\alpha(1-r_{\rm out})}\right] \nonumber \\
1-r_{\rm in} &=& \left[1-P^{\rm out}(0)\right]^2 
= \left[ 1 - \left[1 - \alpha (1-r_{\rm in})\right]
e^{-2\alpha(1-r_{\rm out})} \right]^{2}
\end{eqnarray}
As anticipated, for $\alpha < \alpha_d := 0.883867$ the only solution
is $r_{\rm out}=r_{\rm in} = 1$, whereas above $\alpha_d$ a non
trivial solution appears.

It is easy to show that Eqs.(\ref{eq:transition_alpha}) are equivalent
to the leaf removal ones (and indeed they give the same
threshold). Indeed, the probability for an incoming (into a node
regulated by a given constraint) probability distribution of biases
not to be trivial must be equal to the probability that both input
nodes are not leaves in the pure XOR case; while the probability for
an outcoming distribution of biases not to be trivial will be the
probability that the remaining input node and the output node are not
leaves. This translates immediately into relations

\begin{eqnarray}
\label{corrispondenze}
1 - r_{\rm out} &=& (1 - t_{\rm out}) (1 - t_{\rm in}) \nonumber \\
1 - r_{\rm in} &=& (1 - t_{\rm in})^2 
\end{eqnarray} 

Substituting (\ref{corrispondenze}) into (\ref{eq:transition_alpha}) one
immediately retrieves LR relations.

In the $y \rightarrow \infty $ limit the complexity is given by
\cite{PM1, PM2, MPZ}
\begin{equation}
\Sigma = -y \Phi(y) .
\end{equation}
The free energy is given by $\Phi= \Phi_1 - 2 \alpha \Phi_2 $ with:
\begin{eqnarray}
\label{free_energy}
\Phi_{1} &=& -\frac{1}{y} \left \langle \ln \left(\int Q^{\rm in}(u)^{b}
du \prod_{l=1}^{k} du_{l} Q^{\rm out}(u_l)  e^{y\left(|\sum_{a=1}^k u_{a}
    + b u | -y \sum_{a=1}^k |u_a| -
    |b u |\right)} \right) \right\rangle_{k,b,\{\eta\}} \nonumber \\
\Phi_2 &=& -\frac{1}{y} \left\langle \ln \left( \int dh P^{\rm in}(h) 
      du Q^{\rm in}(u) e^{(-y |u| -y |h| +y |u + h |)}\right) 
\right\rangle_{k,b,\{\eta\}}
\end{eqnarray}
where the averages are taken with respect to the Poissonian
distribution of $k$ defined in Eq.~(\ref{eq:degdistr}), and with respect
to $\rho^{\rm out}(\eta)$ and $\rho^{\rm in}(\eta)$.

\begin{figure}[htb]
\vspace{0.2cm}
\begin{center}
\includegraphics[width=0.65\columnwidth]{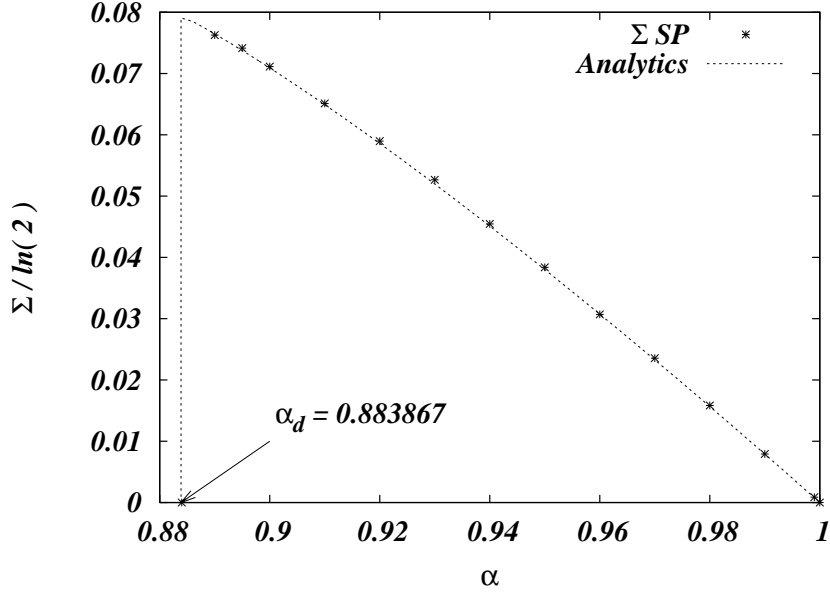}
\end{center}
\caption{Complexity $\Sigma$ vs
  $\alpha$. Dots correspond to the average value obtained by running
  the SP equations on 100 samples of $N=100 000$. Errors are of the size of
  points. The continuous line is the analytic result.}
\label{fig:complexity}
\end{figure}

Following the standard computation outlined in App.~\ref{app:sigma}
we obtain for the complexity of ground states:
\begin{equation}
\label{eq:Sigma}
\Sigma = \lim_{y \rightarrow \infty} y \Phi( y ) = 
\ln 2 \left[ 1 - 2 \alpha(1-t_{\rm out}) -
e^{-2\alpha(1-t_{\rm out})} (1-\alpha (1-t_{\rm in}))\right]
\end{equation}
where $t_{i,o}$ are solutions of Eq.~(\ref{eq:transition_alpha}).  In
Fig.~\ref{fig:complexity} we show the complexity profile in the
clustering region $0.883867 <\alpha< 1$, for the pure XOR model
($x=1$): At $\alpha = 0.883867$ it jumps discontinuously to a non-zero
value and than decreases continuously with growing $\alpha$, until it
vanishes at $alpha=1$. Note that the fixed-point entropy, i.e. the
normalized logarithm of the number of fixed points, is always given by
$(1-\alpha)\ln 2$ as calculated using the BP algorithm, and thus does
not show any discontinuity at the clustering transition. The
observation that the complexity in Fig.~\ref{fig:complexity} is
smaller than this entropy demonstrates that each cluster itself is
exponentially large.

The general case, where canalising and non-canalising function ($x\neq
1$) are taken into account simultaneously, cannot be solved completely
analytically. Nevertheless it is possible to iterate Eq.~(\ref{eq:sp})
numerically on single samples, and give an estimate of the full phase
diagram ($\alpha, x$) including the onset of the clustered phase.

\begin{figure}[htb]
\vspace{0.2cm}
\begin{center}
\includegraphics[width=0.65\columnwidth]{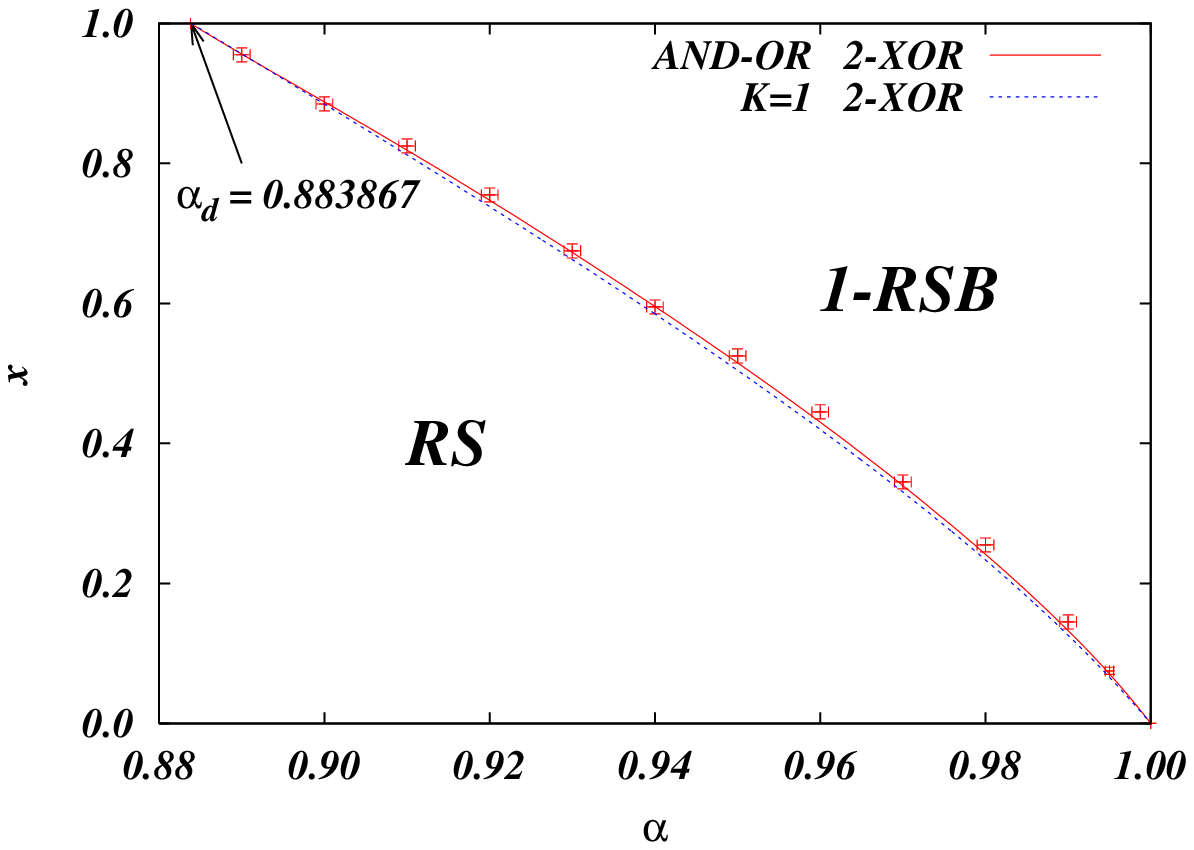}
\end{center}
\caption{Phase diagram $\alpha,x$ of the clustered (1RSB) and
  non-clustered (RS) regions. The red continuous line is a guide to eye
  joining in a smooth way the experimental points. It represents the
  phase boundary of the mix canalising/non canalising $K=2$
  functions. The blue dotted line is the analytic phase boundary of a
  mixture of $K=2$ XOR family and K=1 functions (see second group in
  Tab.~\ref{tab:k2fun}).}
\label{fig:phase_digram}
\end{figure}

In Fig.~\ref{fig:phase_digram} we show the phase diagram $\alpha,x$ of
the clustered (1-RSB) and non-clustered (RS) regions, obtained by
scanning vertical slices ({\em i.e.} at fixed $\alpha$) of the
parameter space. For each point we have analysed 100 samples of
$N=100\ 000$. The $x$-error is estimated by the interval ranging from
the point where all the 100 samples show zero complexity, to the point
where all of the samples have finite complexity. For the
$\alpha$-error we have used the same technique scanning horizontally
at $x$ fixed to the corresponding measured point.

To go beyond the case of real $K=2$ functions ({\it i.e.}~AND-OR and
XOR class functions), we consider first a mixture of $K=2$ non canalising
functions with $K=1$ functions, intended as $K=2$ functions depending
only on one variable (see second group in Tab.~\ref{tab:k2fun}). Here
the combinatorial analysis can be done analytically following closely
the techniques presented for the pure XOR ($x=1$) case. In this
particular mixed case Eq.~(\ref{eq:transition_alpha}) becomes:
\begin{eqnarray}
\label{eq:transition_alpha_x}
r_{\rm out}&=&  1 - x \left[1-\alpha
(1-r_{\rm in})\right] \left[ 1 - e^{-(1+x)\alpha(1-r_{\rm out})} \right]^2  - 
( 1 - x ) \left[1 - e^{-(1+x)\alpha(1-r_{\rm out})}\right]
\nonumber\\
r_{\rm in} &=&  1 -  
x \left[1-\alpha(1-r_{\rm in}) e^{-(1+x)\alpha(1-r_{\rm out})} \right]^2 
- (1- x) 
\left[(1-\alpha)(1-r_{\rm in}) e^{-(1+x)\alpha (1-r_{\rm out})}\right]
\end{eqnarray}
where $x$ is now the fraction of $K=1$ functions. Again, as in the
case of the pure XOR model, at each value of $x$ one can find the
corresponding value of $\alpha$ where both $t_{\rm in},t_{\rm
out}$ are non-trivial. This transition line is the blue dotted line
displayed in Fig.~\ref{fig:phase_digram}. Surprisingly enough, it lies
very close to the AND-OR 2-XOR transition line, but a precise measure
performed at $\alpha = 0.95$ exclude that the two transition lines are
the same.

\begin{figure}[htb]
\vspace{0.2cm}
\begin{center}
\includegraphics[width=0.65\columnwidth]{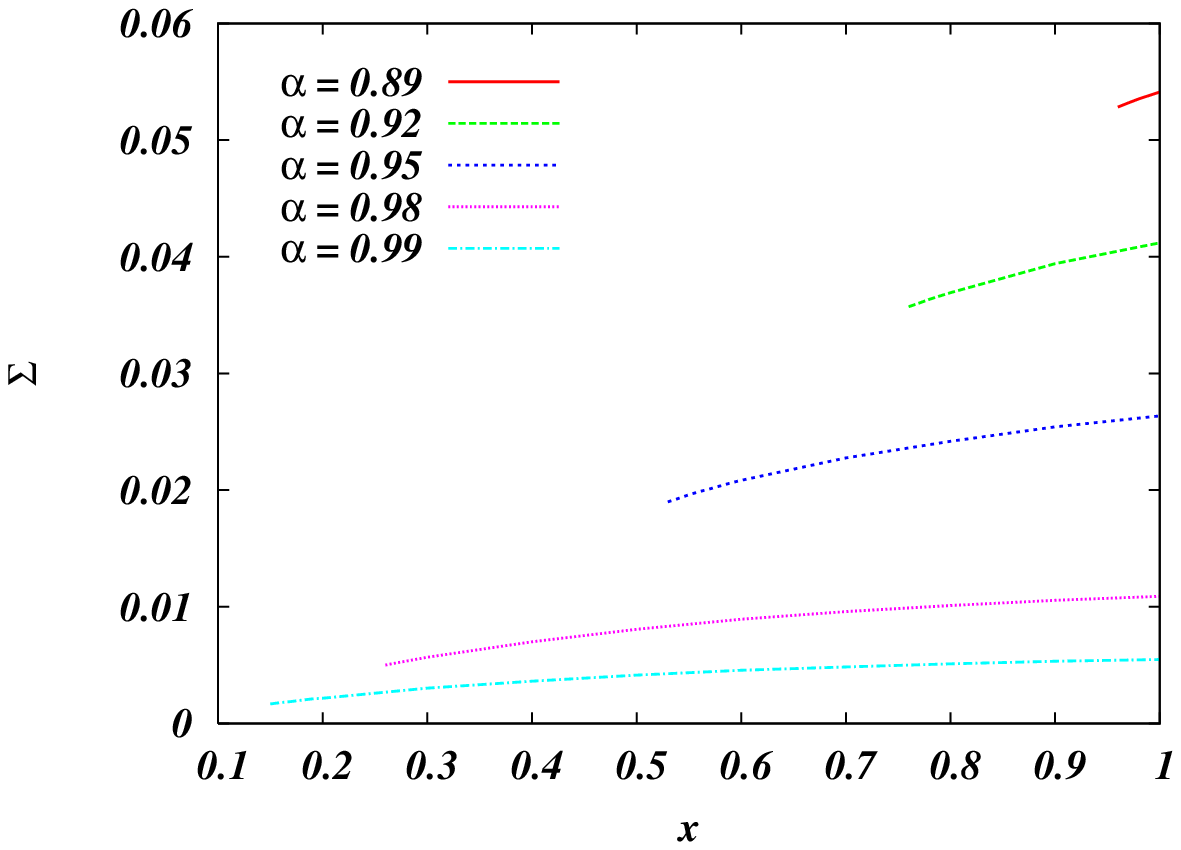}
\end{center}
\caption{Complexity $\Sigma$ vs. concentration of canalising functions
  $x$ at different values of $\alpha$ inside the clustered phase.}
\label{fig:sigmavsx}
\end{figure}

The number of clusters is increasing with $x$ and reaches the maximum,
at any $\alpha \in [\alpha_d, 1]$, exactly in the case of the pure XOR
model at $x=1$. This result is displayed in Fig.~\ref{fig:sigmavsx}
where we plot the complexity $\Sigma$ vs. $x$ at different values of
$\alpha$.  It is interesting also to note that the curves do not
depend on the actual relative concentration of representative of the
AND-OR family, but only on the ratio $x$ between canalising and non
canalising functions. As a limiting case, we have chosen a mixture of
XOR and only direct AND functions (column 9 of Tab.~\ref{tab:k2fun}),
and the result is the same obtained with a flat distribution on the
AND-OR family. The only difference in the latter case is that the
system become magnetised as a result of the lack of symmetry between
$0$ and $1$ which is present in the flat ensemble of canalising
functions. Note that also in all these cases the entropy is given by
the BP value $(1-\alpha)\ln 2$, and remains completely analytic as a
function of $\alpha$ and of the composition of Boolean functions.

\section{Beyond $K=2$} 
\label{sec:K34}

As already mentioned, the restriction to $K=2$ functions is a purely
technical one: the relatively small number of $2^{2^2}=16$ functions
allows a complete characterisation into the four discussed classes,
and it is possible to exhaustively analyse these classes together with
arbitrary mixtures. The number of functions is, however, growing
super-exponentially as $2^{2^K}$ with the input size $K$, making this
exhaustive strategy intractable already for very small $K$. The
general case of $K>2$ needs, in addition, an extension of the
classification into canalising and non-canalising functions. A natural
extension based on the symmetries of Boolean functions are the
so-called NPN functions: equivalence classes are closed under negation
of inputs, permutation of inputs, and negation of the output. It is
very easy to check that this classification reproduces exactly the
four $K=2$ classes discussed before: we simply take one member of the
class, and apply the NPN transformations to reach every other element
in the class, and none outside. To give an example, the AND function
can be transformed into the OR function by negating both inputs and
the output. NPN equivalence classes are also a natural extension of
the classification scheme from $K=2$ to $K>2$ in the sense that the
clustering phenomenon does not depend on the relative appearance of
functions inside each class, but just on the relative appearance of
different classes.

\subsection{$K=3$}
\begin{figure}[htb]
\vspace{0.2cm}
\begin{center}
\includegraphics[width=0.6\columnwidth]{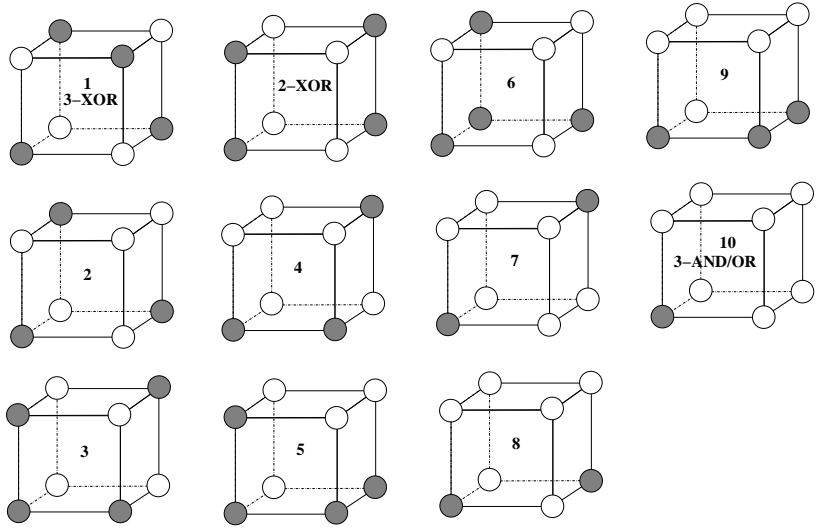}
\end{center}
\caption{Function representatives of the 10 true $K=3$ classes plus
pure $K=2$ XOR. Function are organised in decreasing replica symmetry
breaking order.}
\label{fig:3-cubes}
\end{figure}

In the $K=3$ case, the $2^{2^3} = 256$ functions can still be
classified completely: there are 14 classes, 4 of them have already
been discussed in the context of $K=2$, and only 10 lead to true $K=3$
functions.  Each of the 256 functions can be pictorially represented
on a unit cube (a $K$-cube for generic $K$): The Cartesian position of
any vertex of the cube gives the $K$ input bits, and the output of the
function is represented by a colouring of all vertices: a filled vertex
corresponds to the value $1$ of the output variable, an empty vertex
corresponds to the case where the value $0$ is the satisfying one.
Out of one such function, the NPN equivalence class can be constructed
via the following transformations: exchange of two parallel planes (=
negation of the corresponding input variable), permutations of the
Cartesian coordinates (= permutation of input variables) and inversion
of all vertex colours (= negation of the output). Representatives of
all actual $K=3$ classes are shown in Fig.~\ref{fig:3-cubes}, together
with the 2-XOR function. Note that the independence of a function on
one of the variables can be easily seen in this representation by two
parallel planes carrying identical colours.

For each $K$ there are in total
\begin{equation}
{\cal N}_K = \sum_{t=0}^K \genfrac{(}{)}{0pt}{}{K}{t}
 (-1)^t 2^{2^{K-t}}
\end{equation}
true $K$-functions, leading to ${\cal N}_3 = 218$.  The number of boolean
functions present in each type is reported in Tab.~\ref{tab:k3fun}. 

\begin{table}[ht]
\begin{tabular}{|l|c|c|c|}
\hline
Class & Degeneracy & ${\cal N}_{col}$ & Sub-nodes nature \\
\hline
{\bf 1} ($3-\bigoplus$) & 2 & 0 & $2-\bigoplus$ on all planes \\
{\bf 2} & 24 & 3 & $2-\bigoplus$ on 3 planes. 
$2-\bigvee \bigwedge$ on others  \\
{\bf 3} & 24 & 4 & $2-\bigoplus$ on 3 planes, 
$2-\bigvee \bigwedge$ + $K=1$ on others \\
$2-\bigoplus$ & 6 & 4 & $2-\bigoplus$ on 2 planes, $K=1$ on others  \\
{\bf 4} & 48 & 5 & $2-\bigoplus$ on 1 plane  \\  
{\bf 5} & 24 & 6 & $2-\bigvee \bigwedge$ on all planes  \\
{\bf 6} & 8 & 6 & $2-\bigvee \bigwedge$ on 4 planes, $K=1$ on others \\
{\bf 7} & 8  & 5 & $2-\bigvee \bigwedge$ on all planes \\
{\bf 8} & 24  & 6 & $2-\bigoplus$ on 1 plane, but $K=0$ on another, 
$2-\bigvee \bigwedge$ on others  \\
{\bf 9} & 48 & 7 & $2-\bigvee \bigwedge$ on 3 planes, 
$K=0,1$ on others \\
{\bf 10} ($3-\bigvee \bigwedge$) & 16 & 9 & 
$2-\bigvee \bigwedge$ on 3 planes, $K=0$ on others  \\
\hline
\end{tabular}
\caption{Classification of the $K=3$ boolean functions.}
\label{tab:k3fun}
\end{table}

\begin{figure}
\vspace{0.2cm}
\begin{center}
\includegraphics[width=0.6\columnwidth,angle=270]{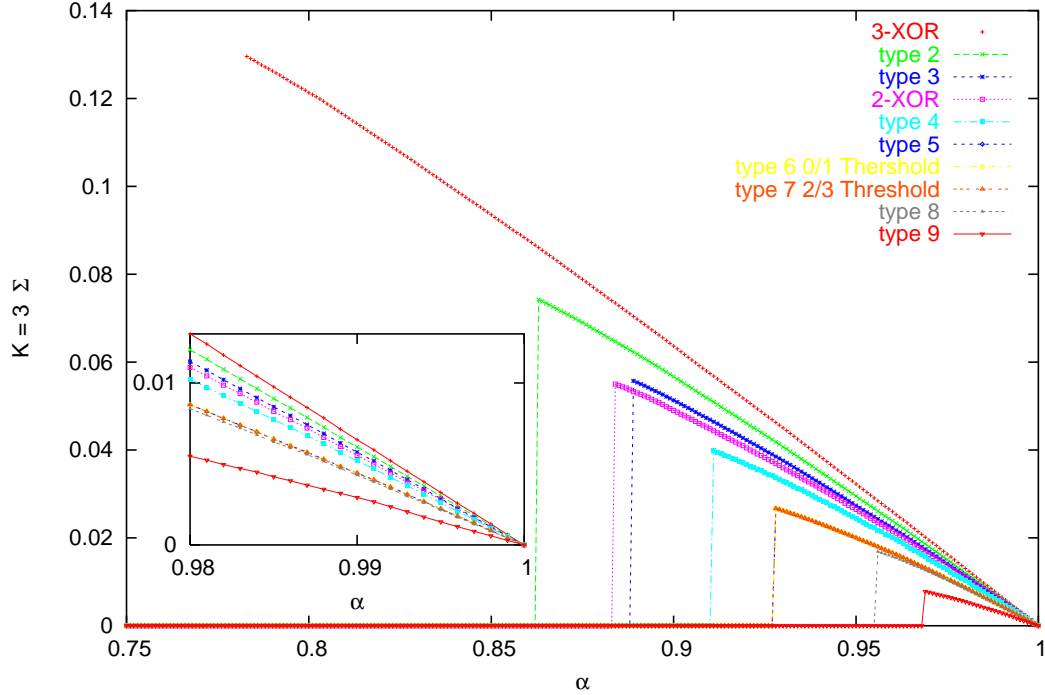}
\end{center}
\caption{Ground state complexity curves vs $\alpha$. Inset:
enlargement of the region close to  $\alpha = 1$: no crossing is
observed.}
\label{fig:3-sigma}
\end{figure}

Function are organised in decreasing replica symmetry breaking order:
function types that undergo the ground-state clustering transition
earlier in $\alpha$ are drawn first. Complexity curves for BN
including purely functions from one class are shown in
Fig.~\ref{fig:3-sigma}. As for $K=2$, each curve is averaged over
$100$ samples at $N = 10^5$. Within one NPN class, functions where
chosen at random with uniform probability. We tested that sample to
sample fluctuations are small (smaller than the points size) already
for $N = 10^4$. Finite size effect become negligible beyond $N =
10^5$.

Generally, for function types that break replica symmetry earlier, the
complexity is higher. This rule has an exception for types $3$ (see
Fig.~\ref{fig:3-sigma}). Complexity curves for each function type do
no intersect below the critical point at $\alpha = 1$, where all
complexities vanish simultaneously. Function types $5$, $6$ and $7$
seem to have a complexity degeneracy along the whole $\Sigma$ curve.
The degeneracy is resolved for type $7$ with tests at $N = 10^6$, but
persist for the other two types. We do not expect this degeneracy to
be exact, but numerical values on an average of $100$ $N=10^5$ samples
are not distinguishable, as well as on a test at $N = 10^6$.

It is also possible to rank function types according to their
``distance'' from the pure XOR node. There are, however, many
non-rigorous measures of distance. One intuitive example is the number
${\cal N}_{col}$ of cube links connecting two vertices of identical
colour. Another possibility is looking at the XOR-like nature of the
node on sub-planes, {\em i.e.}~with one fixed input variable. Both
indicators are shown in Tab.~\ref{tab:k3fun}.

$3-AND-OR$ function types are the only ones that do not show a
complexity transition. They are also the completely canalising nodes,
{\em i.e.}~canalising in every variable.

\begin{figure}
\vspace{0.2cm}
\begin{center}
\includegraphics[width=0.6\columnwidth]{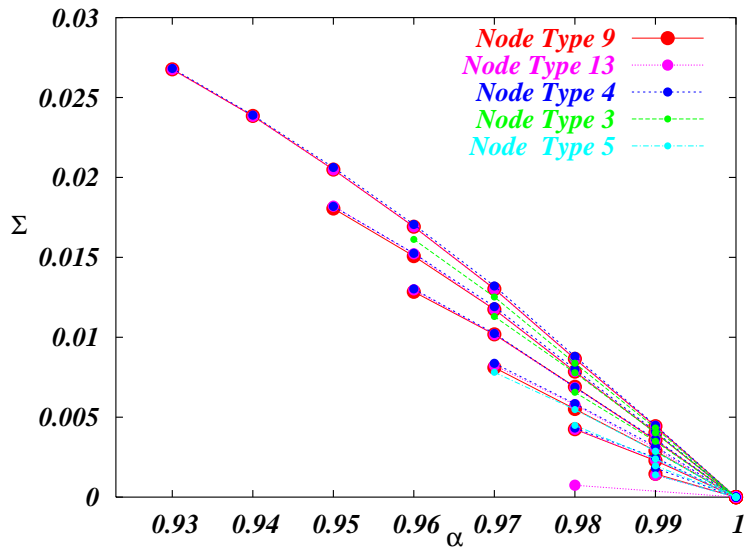}
\end{center}
\caption{Ground state complexity curves vs $\alpha$ for nodes types 9
(red), 13 (pink), 4 (dark blue), 3 (green), 5 (light blue) for different
increasing values of $F_{\vee, \wedge}$. Type 9,13,4: $F_{\vee, \wedge} =
0.0,0.1,0.2,0.3,0.5$. Type 3,5:$F_{\vee, \wedge} = 0.0,0.1,0.2,0.3$.}
\label{fig:sigma-mixed}
\end{figure}

An exhaustive test of all possible mixtures of function types goes
beyond the scope of this work. In order to test whether any small
fraction of any RSB-responsible function type leads to an (albeit
small) RSB transition close to $\alpha=1$, we calculated the
complexity for the 5 nodes types for which clustering is less
pronounced, mixing each type with a given fraction $F_{\vee, \wedge}$
of pure AND-OR nodes. As expected, the transition point shifts
continuously toward $\alpha=1$ for each case, and the complexity
curve lowers, nevertheless without disappearing for $\alpha<1$.  We
conjecture therefore a complex phase to be always present in the large
$N$ limit, besides in the single boundary case of $F_{\vee, \wedge} =
1$. Some complexity curves are shown in Fig.~\ref{fig:sigma-mixed}.

\subsection{$K\geq 4$}

For $K\geq 4$, we have done some non-exhaustive numerical checks, and
extended the analytical calculation of the pure XOR-type model.  The main
results are:
\begin{itemize}
\item[(i)] The minimum of the clustering point $\alpha_d$ over all
combinations of functional classes is always given by the pure $K$-XOR
class (the only one being completely analytically accessible). We find
$\alpha_d=0.782,\ 0.699$ for $K=3,\ 4$ and $\alpha_d=(\ln K + \ln\ln K
+ \ln\ln\ln K+...)/K$ for $K\gg 1$. Note that the range of the complex
phase becomes larger with $K$, and asymptotically approaches zero.
\item[(ii)] We conjecture that, for fixed $K$, the only class not
leading to clustering is the pure $K$-AND-OR one. This has been
checked explicitly for $K=2,3$, and non-exhaustively also for
$K=4$. This class becomes, however, both combinatorially and
biologically less important for higher $K$, where threshold-like
functions are expected to be more relevant.
\end{itemize}

The non-exhaustive numerical checks for the $K=4$ case where done on
threshold-like functions (selected due to the conjectured biological
relevance at high $K$ values), and on a subset of 10 NPN classes
that are expected to undergo a less pronounced clustering transition
on the basis of the $K=3$ insight. We always found a transition point
close to but distinct from $\alpha = 1$.

Moreover, it is clear from the nature of the nodes (see again
Fig.~(\ref{fig:3-cubes})) that neither local nor global
paramagneticity of the system (as opposed to complete XOR nodes) needs
to be preserved in order to observe clustering. It is therefore
possible to build a whole zoology of networks that leave great freedom
on the numerical value of the number of solution clusters AND of
the fraction of variables fixed in a preferred direction (active or
inactive from the point of view of boolean control).

\section{Applications to biological networks}
\label{sec:appl}
After having analysed in great detail the sudden emergence of a
computational core and the number and organisation of fixed points in
random Boolean networks, we are now applying some of the ideas to
known gene-regulatory networks. The methods used in the analysis of
ensembles of random Boolean networks, namely percolation arguments and
the cavity method, can be translated into algorithms acting on single
graphs. This is obvious for the determination of the core since its
very definition is iterative, but it is also true for the cavity
method which can be reformulated in terms of message-passing
techniques, including belief propagation and survey propagation as
given in Eqs.~(\ref{eq:bp}) and (\ref{eq:sp}). The results below have
to be understood as an illustration for potential applications of our
approach to other large-scale networks.

\subsection{The core of the gene-regulatory network of baker's yeast}
\label{sec:yeast}
One of the largest-scale genome-wide GRN known so far is the one of
baker's yeast {\it S. cerevisi\ae}, for which we used data from Uri
Alon's lab \cite{Alon}. For this regulatory network, only the
topological structure is known, {\it i.e.}~a list of transcription factors
to regulated genes is given, but the specific combinatorial control of
the expression level depending on the existence and/or concentration
of the transcription factors is unknown.

Therefore, we were able to apply only the strictly topological part of
the PER and LR removal procedures, the function-specific steps for
canalising functions in PER and for non-canalising ones in LR were
omitted. 

\begin{figure}
\vspace{0.2cm}
\begin{center}
\includegraphics[width=0.3\columnwidth]{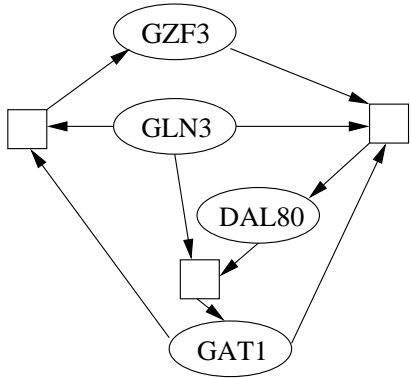}
\end{center}
\caption{The computational core of the yeast GRN consists of the four
genes which are responsible for the regulation of uptake and
degradation of nitrogen sources. All of them are transcription
factors, regulating (in part jointly) the expression of 26 further
genes which all code for functional proteins.}
\label{fig:yeast}
\end{figure} 

The computational core found in this way is presented in
Fig.~\ref{fig:yeast}. Even if it consists only of four genes, this
result is highly interesting. The present genes are exactly those
four genes in the network which are responsible for nitrogen control
in yeast. Nitrogen is, however, the most important mineral nutrient
for yeast. If it is present in abundance, yeast cells live and
proliferate independently from each other. Under nitrogen starvation,
yeast cells perform some cooperative action: Under cell division they
grow into multi-cellular filaments, the so-called {\it pseudohyph\ae},
which are capable to invade new, nutrient rich regions of the surrounding
medium. The main mechanism hereby is nitrogen catabolite repression
(NCR) \cite{Cooper82,Wiame85}, which down-regulates proteins
responsible for using secondary nitrogen sources in the presence of
the ``preferred'' sources. The behaviour of this gene core was studied
in detail only very recently \cite{Boczko05} on the basis of
differential equations.

\subsection{The network of embryonic development in fruit fly}

The segment-polarity genes in fruit fly {\em drosophila melanogaster}
are the last step of a cascade of regulatory interactions responsible
for the onset of dorso-ventral segmentation during the embryonic stage
of development. The stable maintenance of segment-polarity gene
expression is crucial in the development and stability of the
parasegmental segment furrows \cite{Wolpert98,Albert}.

A detailed Boolean model, being able to capture wild type and various
mutant gene expression patterns as fixed points of the network, has
been recently proposed by Albert and Othmer \cite{Albert}. The model
consists of a network of 60 variables and 56 function nodes as a
representation of a single parasegment of four cells, each one composed
of 15 Boolean variables and 14 function nodes.

The network is therefore modular and each module can be described in terms of
the following set of boolean equations:
\begin{eqnarray}
\label{eq:stable} 
SLP_i&=&\left\{\begin{array}{lllll}
0 &\mbox{if}& i\mbox{} \mod4=1 &\mbox{or}& i\mbox{} \mod4=2\\
1 &\mbox{if}& i\mbox{} \mod4=3 &\mbox{or}& i\mbox{} \mod4=0\\
\end{array}\right.\nonumber\\ 
wg_i&=&(CIA_i \mbox{ and } SLP_i \mbox{ and not } CIR_i) \mbox{ or }  
[wg_i \mbox{ and } (CIA_i \mbox{ or } SLP_i ) \mbox{ and not }
CIR_i]\nonumber\\ 
WG_i&=&wg_i\nonumber\\  
en_i&=&(WG_{i-1} \mbox{ or } WG_{i+1}) \mbox{ and not } SLP_i\nonumber\\ 
EN_i&=&en_i\nonumber\\ 
hh_i&=&EN_i \mbox { and not } CIR_i\nonumber\\
HH_i&=&hh_i\nonumber\\ 
ptc_i&=&CIA_i \mbox{ and  not } EN_i \mbox{ and not } CIR_i \\ 
PTC_i&=&ptc_i \mbox { or } (PTC_i \mbox { and not } HH_{i-1} \mbox{ and not } 
HH_{i+1})\nonumber\\ 
PH_i&=&PTC_i \mbox { and } (HH_{i-1} \mbox { or } HH_{i+1})\nonumber\\ 
SMO_i&=&\mbox { not } PTC_i \mbox { or } HH_{i-1} \mbox { or }
HH_{i+1}\nonumber\\
ci_i&=&\mbox{not } EN_i\nonumber\\ 
CI_i&=&ci_i\nonumber\\ 
CIA_i&=&CI_i \mbox { and } ( SMO_i \mbox { or } hh_{i-1} \mbox { or } 
hh_{i+1})\nonumber\\ 
CIR_i&=&CI_i \mbox { and not } SMO_i \mbox { and not } hh_{i-1} 
\mbox { and not }hh_{i+1}\nonumber 
\end{eqnarray}
where module index $i$ runs from 1 to 4. 

As a first test, we have applied the topological part of the LR+PER
algorithm determining the computational core of the BN defined in
Eqs.~(\ref{eq:stable}). We have found that the whole network, except
for the nodes $\{PTC_i\}_{i=1,\dots,4}$, forms the computational core.
We find ourselves thus in the opposite situation of the yeast network,
where only the small nitrogen regulatory module belongs to the
CC. This difference is even more astonishing since the BN
(\ref{eq:stable}) is only a small sub-network of the full
gene-regulatory mechanism in {\it drosophila}. Our observation
underlines therefore the more complex structure of GRNs of higher
organisms compared to simpler ones.

BN (\ref{eq:stable}) is known to have 10 solutions, among which only
one represents the wild type \cite{Albert}. We have applied belief
propagation to the network. Even if this algorithm is expected to
become exact only on large, locally tree-like networks, its prediction
of about 12 fixed points is rather accurate compared to the exact
result. More interesting is, however, the result of {\it in silico
gene-knockout experiments}: Fixing single variables to a value not
occurring in any of the ten fixed points, BP immediately predicts a
negative entropy. Seen the discrete nature of the Boolean variables,
this result has to be interpreted in terms of non-existing fixed
points. BP is thus able to identify {\it essential variables}. This
strategy can be easily extended to {\it in silico} experiments
restricting more than one variable and thus allows to generate
valuable hypotheses and suggestions for real experiments. 

In order to recover fixed points, one has to setup {\em ad hoc}
strategies for fixing variables. BP can be exploited in order to
iteratively generate solutions of the problem using a {\em decimation}
technique \cite{MPZ, Libro_Martin,BMZ}. The idea is the following: run
BP, fix the most biased Boolean variable ({\em i.e.} ~the variable
having the highest probability to be 0 or 1) to its more probable
Boolean value, and iterate the previous 2 steps until all spins are
fixed. Surprisingly enough, the decimation process based on BP leads
to the wild-type expression pattern - which is also the one having the
largest basin of attraction from the dynamical point of view.

We have also checked the robustness of the fixed points of
Eqs.~(\ref{eq:stable}) with respect to noise. In a Boolean setting,
the latter amounts to fluctuations in the output of the Boolean
functions. We have checked this in the {\em drosophila} network: noise
is introduced by replacing the term $1-E_a (\{x_l\}_{l\in a})/2$ with
the finite-temperature Boltzmann term $\exp[-E_a(\{x_l\}_{l\in
a})/T]$. For not too large formal temperature $T$, we always recover
the correct wild-type solution under decimation. Increasing
continuously the noise, the distance of the BP solution to its
zero-temperature counter part increases slowly. This technique can be
used in order to prune out of the whole set of solutions those that
are most sensitive to perturbations.

Finally, we have also run the SP algorithm on this network, but we did
not find any sign of clustering solutions (even in case we relax the
initial conditions, letting free all variables $SLP_i$ - see the first
of Eq.~(\ref{eq:stable})).

\section{Conclusion and outlook}
\label{sec:concl}

\begin{figure}[htb]
\vspace{0.4cm}
\begin{center}
\includegraphics[width=0.65\columnwidth]{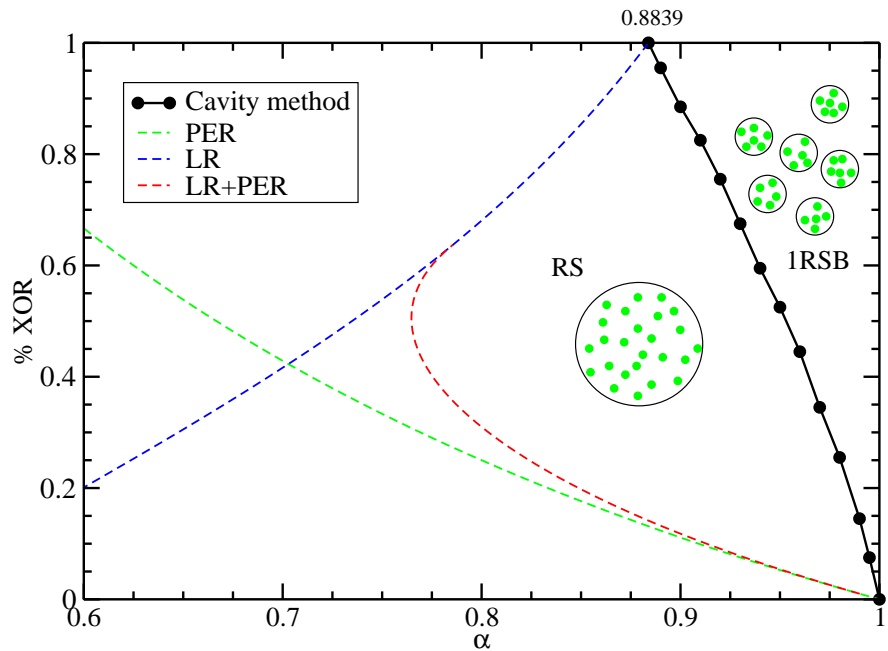}
\end{center}
\caption{Phase diagram for $K=2$: The green line gives the percolation
transition for the PER core, the blue one for the LR core. The red
curve results from the combination of LR and PER, and merges in the
tricritical point (0.785,0.636) with the LR curve. Below this point,
the transition is continuous, above discontinuous in the core
size. The full black line indicates the RSB transition from an
unstructured solution set to a highly clustered one.}
\label{fig:phasediag}
\end{figure}

In this paper, we have studied in detail the structure and
organisation of fixed points in large random Boolean networks. To do
so, we have reformulated the problem in terms of a
constraint-satisfaction problem, and we have applied various
statistical-physics tools ranging from percolation arguments to the
cavity approach from spin-glass theory. The main results of these
approaches are summarised in Fig.~\ref{fig:phasediag}:
\begin{itemize}
\item Networks of small density $\alpha$ of regulatory functions are
dominated by logical cascades and under-constrained variables. At a
certain density, an extensive computational core of the network
emerges suddenly, containing in particular non-trivial feed-back
loops. The existence of such a computational core is a necessary but
not sufficient condition for a non-trivial global behaviour of the
network fixed points.
\item At a higher density, the set of all solutions undergoes a
clustering transition, as schematically represented in
Fig.~\ref{fig:phasediag}: Below the transition line, all fixed points
are collected in one large connected cluster. Above the transition, an
exponential number of these clusters exists, each of them containing
still an exponetial number of solutions. Any two of these clusters
have an extensive Hamming distance. This phenomenon is found to be
robust with respect to the selection of the Boolean functions, and
covers an increasing part of the phase diagram if we go to a higher
number $K$ of inputs to the Boolean functions.
\end{itemize}
All the methods applied to random Boolean networks have also been
formulated as algorithms which can be applied to analyse arbitrary
networks. As a demonstration of possible applications, we have run
these algorithms on two biological gene-regulatory networks, namely
those of yeast and of the segment-polarity genes in fruit fly. We have
first confirmed that the method is able to identify a biologically
sensible computational core of both networks. The message passing
algorithms based on the cavity method ({\it i.e.} belief and survey
propagation) could be applied only to the fruit-fly case where the
Boolean regulation is known: the number of fixed points was predicted
rather accurately. In some {\it in silico gene-knockout experiments},
belief propagation was further on able to identify {\it essential
genes}.

There are, however, various open questions. First, our analysis is
based on ground states of a Hamiltonian counting the number of
violated functions, which is not directly related to any biologically
motivated dynamics. The {\it dynamical accessibility} of fixed points
remains an open challenge which will be addressed in a future
project. Moreover, noise sources present in real cases may force the
network to quasi-stationary points close to the fixed ones (some
regulation might not always function properly, without affecting the
overall state of a cell). In this view, the study of the organisation
and accessibility of meta-stable states in the region of low
complexity and in the case of fixed external inputs, together with
their relation with quasi-stationary points, might be relevant.

A second major direction for future research should be the application
to more complex biological cases, which is expected to give results
going beyond the simple test cases reported in this work. However,
very few genome-wide networks are available so far. In particular, for
multi-cellular organisms only small functional modules for
well-described functions are known. Large-scale networks known for
yeast \cite{GuBoBoKe,Alon} and {\it E. coli} \cite{ShMiAl} contain
only the topological structure, not an extensive description of the
regulatory functions. It is thus highly interesting to infer gene
regulation networks from experimentally easily accessible
high-throughput experiments, as {\it e.g.}~given by genome-wide
expression profiles \cite{600Genes}. This inverse problem can be
treated with tools similar to the ones used in the present analysis.

{\em Acknowledgments} -- We deeply thank Alfredo Braunstein for
allowing us to use his SP implementation, and Fran{\c c}ois K\'ep\`es
for many interesting discussions. This work was supported by EVERGROW
(IST integrated project No. 1935, 6th EU Framework Programme).

\appendix

\section{On the computational core: Combining LR and PER}
\label{app:lr+per}
In this appendix we give some technical details on how to determine
the action of PER on the LR core, i.e. on how to determine the
computational core of a random Boolean network with given external
condition.

\subsection{The degree distribution of the LR core}

To start with, we need the degree distribution for the LR core. In
this context we characterise every variable by the 4-tuple $(d_{\rm
in}^c,d_{\rm in}^n,d_{\rm out}^c,d_{\rm out}^n)$ counting the number
of in-links coming from canalising and non-canalising functions and
the number of out-links going to canalising and non-canalising
functions. Note that the total in-degree $d_{\rm in}^c+d_{\rm in}^n$ of
a variable is either zero or one.  We have to distinguish the
following cases:
\begin{enumerate}
\item The variable is neither leaf nor isolated, $d_{\rm in}^c+d_{\rm
in}^n+d_{\rm out}^c+d_{\rm out}^n\geq 2$. The fraction of variables
having these degree values is given by the generalised Poissonian
\begin{eqnarray}
P_{LR}(d_{\rm in}^c,d_{\rm in}^n,d_{\rm out}^c,d_{\rm out}^n) 
&=& \left[ \{ 1-\alpha x (1-t_{\rm in})^2
-\alpha (1-x) (1-t_{\rm in}^2) \} \delta_{d_{\rm in}^c,0}
\delta_{d_{\rm in}^n,0}
+\alpha x (1-t_{\rm in})^2 \delta_{d_{\rm in}^c,0}\delta_{d_{\rm in}^n,1}
\right.\nonumber\\
&& \left.
+\alpha (1-x) (1-t_{\rm in}^2) \delta_{d_{\rm in}^c,1}\delta_{d_{\rm in}^n,0} 
\right]
\ \ \times \nonumber\\
&&\times e^{-2\alpha(1-t_{\rm out})(1-xt_{\rm in})}\ 
\frac{[2\alpha(1-x)(1-t_{\rm out})]^{d_{\rm out}^c}[2\alpha x (1-t_{\rm in})
(1-t_{\rm out})]^{d_{\rm out}^n}}
{d_{\rm out}^c! d_{\rm out}^n!}
\label{eq:plr1}
\end{eqnarray}
There we have used, e.g., the fact that an in-link from a canalising
function survives if not both inputs of the function are
simultaneously leaves at a certain point of the removal procedure,
which leads to the factor $(1-t_{\rm in}^2)$ in the $(\delta_{d_{\rm
in}^c,1}\delta_{d_{\rm in}^n,0})$-term, and similar arguments for
other in- and out-links.
\item The variable is a leaf, $d_{\rm in}^c+d_{\rm in}^n+d_{\rm
out}^c+d_{\rm out}^n=1$. According to the removal procedure, this is
possible only if the existing link is an input to an AND/OR-type
function:
\begin{eqnarray}
P_{LR}(1,0,0,0) &=& 0 \nonumber\\
P_{LR}(0,1,0,0) &=& 0 \nonumber\\
P_{LR}(0,0,1,0) &=&   e^{-2\alpha(1-t_{\rm out})(1-xt_{\rm in})} \ 
\{ 1-\alpha x (1-t_{\rm in})^2 -\alpha (1-x) (1-t_{\rm in}^2) \}\ 
 2\alpha(1-x)(1-t_{\rm out}) (1-t_{\rm in})
\nonumber\\
P_{LR}(0,0,0,1) &=& 0 
\end{eqnarray}
The non-trivial term is a product of the Poissonian contribution in
Eq.~(\ref{eq:plr1}) and the probability $(1-t_{\rm in})$ that the
other input to the adjacent function is not a leaf. The contrary would
lead to removal of the Boolean function.
\item The variable is isolated, $d_{\rm in}^c+d_{\rm in}^n+d_{\rm
out}^c+d_{\rm out}^n=0$. Note that in LR we remove only the functions,
and the number $N$ of vertices remains unchanged. The isolated
vertices thus collect all originally isolated vertices and all those,
which were leaves leading to a removal step. We consequently find
\begin{eqnarray}
P_{LR}(0,0,0,0) &=&   e^{-2\alpha(1-t_{\rm out})(1-xt_{\rm in})} \ 
\{ 1-\alpha x (1-t_{\rm in})^2 -\alpha (1-x) (1-t_{\rm in}^2) \}\times
\nonumber\\ && \times
[1+2\alpha x(1-t_{\rm in})(1-t_{\rm out}) 2\alpha(1-x)(1-t_{\rm out}) 
t_{\rm in} ]\nonumber\\
&&+e^{-2\alpha(1-t_{\rm out})(1-xt_{\rm in})} \  
\{ \alpha x (1-t_{\rm in})^2 +\alpha (1-x) (1-t_{\rm in}^2) \}
\end{eqnarray}
\end{enumerate}

\subsection{PER on the LR core}
\label{sec:PER}
Now we can come to the analysis of PER on the LR core. The propagation
rules were: A XOR-type function can be removed and its output fixed,
if both inputs are fixed.  An AND/OR fixes its output if either both
inputs are fixed, or if exactly one input is fixed, but to its
canalising value. This iterative rule allows us to give a
self-consistent condition for the probability $\pi_c$ (resp. $\pi_n$)
that an AND/OR type (resp. XOR-type) function fixes its output by PER
and can be removed. (Note that we are now speaking of the removal and
not the survival probability.) For $\pi_n$ we find directly
\begin{equation}
\pi_n = \left[ \sum_{d_{\rm in}^c,d_{\rm in}^n,d_{\rm out}^c,
d_{\rm out}^n} \frac {d_{\rm out}^n}{\langle
d_{\rm out}^n\rangle}\ P_{LR}(d_{\rm in}^c,d_{\rm in}^n,d_{\rm out}^c,
d_{\rm out}^n)\ \pi_c^{d_{\rm in}^c} \
\pi_n^{d_{\rm in}^n} \right]^2\ .
\end{equation}
This equation requires a set of explanations: The brackets
$\langle\cdot\rangle$ denote the average over $P_{LR}$. The
distribution $\frac {d_{\rm out}^n}{\langle d_{\rm out}^n\rangle}\
P_{LR}(d_{\rm in}^c,d_{\rm in}^n,d_{\rm out}^c,d_{\rm out}^n)$ thus
describes the probability that an input link to an arbitrary
non-canalising function is coming from a variable of in- and
out-degrees $(d_{\rm in}^c,d_{\rm in}^n,d_{\rm out}^c,d_{\rm
out}^n)$. This function has to be fixed, which appears with
probability one if the function is not regulated, with probability
$\pi_c$ if it is regulated by a canalising function, with $\pi_n$ if
regulated by a non-canalising one. We may unify all three cases in the
single expression $\pi_c^{d_{\rm in}^c} \ \pi_n^{d_{\rm in}^n}$, since
$d_{\rm in}^c+d_{\rm in}^n=0/1$, i.e.~since a variable is never
regulated by more than one Boolean function. For an AND/OR-type
function, we first denote the probability that one input is fixed by
$P^{(in)}$. Using that a variable takes its canalising value with
probability 1/2, this allows us to write
\begin{eqnarray}
\pi_c &=& P^{(in)} P^{(in)} + 2P^{(in)} (1-P^{(in)}) \times \frac 12
\nonumber\\
&=& P^{(in)} \nonumber\\
&=& \sum_{d_{\rm in}^c,d_{\rm in}^n,d_{\rm out}^c,d_{\rm out}^n} 
\frac {d_{\rm out}^c}{\langle
d_{\rm out}^c\rangle}\ P_{LR}(d_{\rm in}^c,d_{\rm in}^n,d_{\rm out}^c,
d_{\rm out}^n)\ \pi_c^{d_{\rm in}^c} \
\pi_n^{d_{\rm in}^n}\ .
\end{eqnarray}
Using the degree distribution $P_{LR}$ determined in the last
subsection, we can easily calculate the $P_{LR}$-averages as almost
complete Poissonian sums
\begin{eqnarray}
\langle d_{\rm out}^n\rangle &=& 2\alpha x (1-t_{\rm out})(1-t_{\rm in}) 
\left[  1-e^{-2\alpha(1-t_{\rm out})(1-xt_{\rm in})} \ 
\{ 1-\alpha x (1-t_{\rm in})^2 -\alpha (1-x) 
(1-t_{\rm in}^2) \} \right]
\nonumber\\
\langle d_{\rm out}^c\rangle &=& 2\alpha (1-x)(1-t_{\rm in}) 
\left[  1-t_{\rm in} e^{-2\alpha(1-t_{\rm out})(1-xt_{\rm in})} \ 
\{ 1-\alpha x (1-t_{\rm in})^2 -\alpha (1-x) 
(1-t_{\rm in}^2) \} \right]\ .
\end{eqnarray}
A similar resummation can be done for the degree sums in the above
expressions for $\pi_{c/n}$, we obtain
\begin{eqnarray}
\pi_n &=& \left[ \frac{\{1-e^{-2\alpha(1-t_{\rm out})(1-xt_{\rm in})}\} 
\{ 1-\alpha x (1-t_{\rm in})^2 -\alpha (1-x) (1-t_{\rm in}^2) \} 
+ \pi_n \alpha x (1-t_{\rm in})^2 + \pi_c \alpha (1-x) (1-t_{\rm in}^2)}{
1-e^{-2\alpha(1-t_{\rm out})(1-xt_{\rm in})} \ 
\{ 1-\alpha x (1-t_{\rm in})^2 -\alpha (1-x) 
(1-t_{\rm in}^2) \}}\right]^2 \nonumber\\
\pi_c &=& \frac{\{1-t_{\rm in} e^{-2\alpha(1-t_{\rm out})(1-xt_{\rm in})}\} 
\{ 1-\alpha x (1-t_{\rm in})^2 -\alpha (1-x) (1-t_{\rm in}^2) \} 
+ \pi_n \alpha x (1-t_{\rm in})^2 + \pi_c \alpha (1-x) (1-t_{\rm in}^2)}{
1-t_{\rm in} e^{-2\alpha(1-t_{\rm out})(1-xt_{\rm in})} \ 
\{ 1-\alpha x (1-t_{\rm in})^2 -\alpha (1-x) 
(1-t_{\rm in}^2) \}}\ .
\end{eqnarray}
These equation can be simplified using the
Eqs.~(\ref{eq:to},\ref{eq:ti}) for the parameters of the LR core,
\begin{eqnarray}
\pi_n &=& \left[ \frac{t_{\rm in}(1-t_{\rm out})+\pi_n\alpha x
(1-t_{\rm in})^2 + \pi_c \alpha (1-x) (1-t_{\rm in}^2)}{t_{\rm
out}(1-t_{\rm in})}\right]^2 \nonumber\\ \pi_c &=& \frac{ t_{\rm
in}(1-t_{\rm out} t_{\rm in})+ \pi_n \alpha x (1-t_{\rm in})^2 + \pi_c
\alpha (1-x) (1-t_{\rm in}^2)}{t_{\rm out}(1-t_{\rm out} t_{\rm in})}\
.
\end{eqnarray}
These equations can be studied to analyse the emergence of a
non-trivial computational core of the random Boolean network, in
particular the continuous transition to an extensive core below the
tricritical point in Fig.~\ref{perc} can be understood by linearising
the first equation in $\pi_n$ and $\pi_c$. Here we do not give the
steps of this feed-forward calculation.

\section {The computation of the complexity $\Sigma$}
\label{app:sigma}
We will derive in some details the computation of the entropy $\Sigma$
at $x=1$. Let us compute first the $\Phi_1$ contribution:
\begin{eqnarray}
\label{free_energy1}
\Phi_1 &=& -\frac{1}{y} \left \langle \ln \left(\int Q^{\rm in}(u)^{b}
du \prod_{l=1}^{k} du_{l} Q^{\rm out}(u_l)  e^{y\left(|\sum_{a=1}^k u_{a}
    + b u | -y \sum_{a=1}^k |u_a| -
    |b u |\right)} \right) \right\rangle_{k,b,\{\eta \}} \nonumber \\
&=&\alpha \sum_{k=0}^\infty e^{-2\alpha}\frac{(2 \alpha)^k}{k!} 
\int \prod_{l=1}^{k} d\eta_l \rho^{\rm out}(\eta_l) d\eta \rho^{\rm in}(\eta)
\ln{\left[ \int \prod_{l=1}^k du_{l} Q^{\rm out}(u_l) du Q^{\rm in}(u)
    e^{\left( y |\sum_{l=1}^k u_{l} + u | -y \sum_{l=1}^k |u_l| - y
      |u|\right)}\right]} + \nonumber \nonumber \\
 && \,\,\,\,\,\,\,\,\,\, + (1-\alpha) \int \prod_{l=1}^{k} d\eta_l
\rho^{\rm out}(\eta_l)\ln{\left[ \int \prod_{l=1}^{k} du_{l} Q^{\rm out}(u_l)
    e^{\left( y |\sum_{a} u_{a}| -y \sum |u_{a}|\right)}\right]}
\nonumber \\
&=& (1-\alpha(1-t_{\rm in})) \sum_{k} e^{-2\alpha}\frac{(2 \alpha)^k}{k!}
\int \prod_{l=1}^{k} d\eta_l \rho^{\rm out}(\eta_l) d\eta \rho^{\rm in}(\eta) 
\ln{\left[2 \prod_l \frac{(1+\eta_l)}{2} \frac{(1+\eta)}{2} 
-\prod_l \eta_l \eta\right]} +  
\nonumber\\
& &\,\,\,\,\,\,\,\,\,\, 
+ \alpha t_{\rm in} \sum_{k} e^{-2\alpha}\frac{(2 \alpha)^k}{k!}
\int \prod_{l=1}^{k} d\eta_l \rho^{\rm out}(\eta_l) 
\ln{\left[2 \prod_l (\frac{(1+\eta_l)}{2}) - \prod_l \eta_l 
\right]}\nonumber\\
&=& \ln 2 (1-\alpha (1-t_{\rm in})) \left( 1 - 2\alpha (1-t_{\rm out}) - 
e^{(-2\alpha(1-t_{\rm out}))}\right) 
- 2\alpha^2 (1-t_{\rm in}) (1-t_{\rm out}) \ln2 
\end{eqnarray}

The $\Phi_2$ contribution is computed along the same lines of
$\Phi_1$, and is given by:
\begin{equation}
\Phi_2 = - \ln(2) (1-t_{\rm in}) \left[1-e^{-2\alpha (1-t_{\rm
out})}\right]
\end{equation}
Putting everything together one obtains for $\Sigma$ the result shown
in Eq.~(\ref{eq:Sigma}).

\end{document}